\documentclass[review]{elsarticle}

\usepackage{lineno}
\usepackage{hyperref}

\usepackage{amssymb,amsbsy,amsmath,amsthm,amsfonts,amscd}
\usepackage{nccmath}
\usepackage{relsize}
\usepackage{xcolor}
\usepackage{graphicx}

\usepackage{subfig}
\usepackage{float}
\usepackage{subcaption}
\usepackage{multirow,tabularx}
\usepackage{enumitem}

\usepackage{algorithm2e} 

\usepackage{algcompatible}

\newcommand{\vectb}{\boldsymbol}
\newcommand{\vect}[1]{\mathbf{#1}}

\newcommand{\etal}{\textit{et al.}}
\newcommand{\derpar}[2]{\dfrac{\partial #1}{\partial #2}}

\journal{Journal of Non-Newtonian Fluid Mechanics}
\biboptions{sort&compress}

\expandafter\def\expandafter\normalsize\expandafter{%
    \normalsize%
    \setlength\abovedisplayskip{4pt}%
    \setlength\belowdisplayskip{8pt}%
    \setlength\abovedisplayshortskip{-6pt}%
    \setlength\belowdisplayshortskip{2pt}%
}

\begin{document}

\begin{frontmatter}

\title{A quasi-analytical solution for ``Carreau-Yasuda-like'' shear-thinning fluids flowing in slightly tapered pipes}

\author[indirizzoTVdicii]{Gianluca Santesarti\corref{mycorrespondingauthor}}
\cortext[mycorrespondingauthor]{Corresponding author}
\ead{santesarti@ing.uniroma2.it}

\author[indirizzoTVdicii]{Michele Marino}
\ead{m.marino@ing.uniroma2.it}

\author[indirizzoGSSI]{Francesco Viola}
\ead{francesco.viola@gssi.it}

\author[indirizzoGSSI,indirizzoTVind,indirizzoNL]{Roberto~Verzicco}
\ead{verzicco@uniroma2.it}

\author[indirizzoTVdicii]{Giuseppe Vairo}
\ead{vairo@ing.uniroma2.it}

\address[indirizzoTVdicii]{University of Rome Tor Vergata, Department of Civil Engineering and Computer Science Engineering, 00133 Rome, Italy}
\address[indirizzoGSSI]{Gran Sasso Science Institute, L'Aquila, 67100, Italy}
\address[indirizzoTVind]{University of Rome Tor Vergata, Department of Industrial Engineering, 00133 Rome, Italy}
\address[indirizzoNL]{Physics of Fluids Group, Max Planck Center for Complex Fluid Dynamics, MESA+ Institute and J. M. Burgers Centre for Fluid Dynamics, University of Twente, P.O. Box 217, 7500AE Enschede, Netherlands}

\begin{abstract}
This paper presents a quasi-analytical framework for ``Carreau-Yasuda-like'' fluids
with a viscosity characterized by two constant plateau at low and high shear rates
connected by a shear-thinning branch, and flowing in slightly tapered pipes.
This setup is common in research and industrial applications since the last century,
by assuming both a Newtonian or a non-Newtonian inelastic behaviour.
Nevertheless, an analytical solution for ``Carreau-Yasuda-like'' fluids is still lacking.
The expressions have been derived by using the order-of-magnitude analysis
and neglecting the inertial terms in the momentum balance equations.
The analytical solutions are employed to an extrusion bioprinting process
as an application example, and verified through numerical procedures.
\end{abstract}


\end{frontmatter}
\nolinenumbers

\section{Introduction}
  Fluids flowing in tapered pipes, like conical ducts, are present in a wide
  range of research and industrial applications such as plastic polymer
  manufacturing \cite{bird1987dynamics, agassant2017polymer}, foodstuffs \cite{rao2010rheology},
  and biomedical applications \cite{conti2022models, sarker2017modeling, akbar2014blood, mandal2005unsteady, forrester1970flow}.
  The mathematical modelling of this problem has been addressed since the last
  century both for Newtonian and non-Newtonian fluids.
  
  Regarding the Newtonian case, Blasius (1910) \cite{blasius1910laminare} investigated 
  the axisymmetric steady flow in channels and tubes with a small exponential 
  divergence to analyse the boundary layer separation phenomenon.
  By means of order-of-magnitude analysis and using the method of the ``successive 
  approximations'' he found a solution for the axial and radial velocity components.
  Then, Forrester and Young (1970) \cite{forrester1970flow} studied the boundary 
  layer separation of blood in vascular diseases, considereing slightly converging
  and diverging vessels (mild stenosis and dilatations, respectively).
  In particular, they found a fourth order polynomial solution but only for the 
  axial velocity component.
  Langlois (1972) \cite{langlois1972creeping} analysed the creeping flow (i.e. 
  by neglecting the inertial terms in the momentum balance quations) in a circular 
  tube with a varying cross section, and presented a solution based on the power 
  series expansion in the tangent of the taper angle and assuming the streamlines 
  as straight lines passing through the cone vertex.
  Kotorynski (1995) \cite{kotorynski1995viscous} provided a solution for both the 
  axial and radial velocity components based on the recursive successive 
  approximations method, as a function of the axial pressure gradient and 
  employing the symbolic manipulation language Maple.
  Then, Sisavath \textit{et al.} (2001) \cite{sisavath2001creeping} improved the 
  study of Forrester and Young (1970) \cite{forrester1970flow};
  they derived an asymptotic solution of the Navier-Stokes equations at low Reynolds 
  numbers which does not account for the wavelength of the channel constriction 
  by adopting the perturbation analysis.

  However, in many advanced applications the working fluids exhibit a complex non-Newtonian
  and nonlinear rheological behaviour \cite{bird1987dynamics}.
  In particular, the class of generalized Newtonian fluids (GNFs), also known as
  viscous inelastic fluids, manifest shear-thinning and visco-plastic effects.
  For these fluids, the actual shear stress is a function of the shear rate
  at the current time, and it can be described via a generalized form of the
  constitutive equation of Newtonian fluids, in which the apparent (or effective)
  viscosity is a nonlinear decreasing function of the shear rate.
  Their rheological response has been modelled in the literature through several
  empirical models,
  such as the power-law model \cite{waele1923viscometry, ostwald1925ueber},
  the Casson model \cite{casson1959rheology},
  the Bingham model \cite{bingham1922fluidity},
  the Herschel-Bulkley model \cite{herschel1926konsistenzmessungen},
  the Cross model \cite{cross1965rheology},
  the Carreau model \cite{carreau1972rheological}
  and the Carreau-Yasuda model \cite{yasuda1979investigation, bird1987dynamics}.

  Several studies on such non-Newtonian shear-thinning fluids flowing in tapered
  pipes have been presented to date.
  Sutterby (1966) \cite{sutterby1966laminar} proposed an alternative rheological model
  to evaluate the flow rate-pressure drop relationship through a numerical method. 
  Oka and Murata (1969) \cite{oka1969theory} provided general integral solutions 
  for the shear stress, velocity and flow rate by neglecting the inertial terms 
  in the momentum balance equations.
  Walawender and Prasassarakich (1976) \cite{walawender1976flow} compared the 
  flow rate-pressure drop relationship of a Casson fluid flowing in conical vessels 
  and equivalent cylindrical vessels. 
  Then, How \textit{et al.} (1987) \cite{how1988comparison} applied the solution 
  presented by Oka and Murata \cite{oka1969theory} to polyacrylamide solutions 
  with viscosity data fitted through the power-law model, in order to study the
  pressure losses of the blood flow in arterial prostheses.
  More recently, Priyadharshini and Ponalagusamy (2015) \cite{priyadharshini2015biorheological} 
  improved the solution of Forrester and Young (1970) regarding the study of vascular 
  diseases by modelling the blood as a Herschel-Bulkley fluid and providing a
  solution for the axial velocity component of the flow. 
  Then, Paneseti \textit{et al.} (2018) \cite{panaseti2018pressure} analysed
  the lubrification flow of a Herschel-Bulkley fluid in a symmetric channel with
  varying width through a semi-analytical approach.
  Next, Fusi \textit{et al.} (2020) \cite{fusi2020flow} provided a semi-analytical
  solution for the flow of a Bingham fluid in a variable radius pipe. 
  However, regarding the Carreau-Yasuda model, even though its wide employement
  in many applications such as plastic manufacturing \cite{mazzanti2016rheological},
  hemodynamics \cite{gijsen1999influence}, bioprinting \cite{sauty2022enabling, chirianni2024development, chirianni2024influence},
  lubricant production \cite{bair2006more}, and food processing \cite{meza2021effect},
  an analytical solution for ``Carreau-Yasuda-like'' fluids has not yet been provided.

  This work presents an approximated analytical solution for ``Carreau-Yasuda-like''
  fluids flowing in slightly tapered axisymmetric pipes.
  The approximation lies in replacing the viscosity rheological response with a
  piecewise approximation, characterized by two constant plateau at low
  and high shear rates connected by a shear-thinning branch.
  The solution is derived in the viscous limit, hence when the inertial convective
  terms in the Navier-Stokes equations are negligible.
  The derived analytical solution is applied to polymer flows used in the
  biomedical application of extrusion bioprinting and verified through numerical 
  solutions. 

\section{Rheological modelling}\label{sec:2 Rheol. models} 
  For incompressible GNFs, the constitutive relationship between the deviatoric
  stress tensor $\vectb{\tau}$ and the strain-rate tensor $\vectb{E}$ reads \cite{bird1987dynamics}:
  \begin{equation}\label{eq:GNF definition}
    \vectb{\tau} \left(\dot{\gamma}\right) = 
    2\mu\left(\dot{\gamma}\right) \vectb{E}\ = \mu \left(\dot{\gamma}\right)\  \left( \nabla\vectb{v}+\nabla^T\vectb{v} \right),
  \end{equation}
  where $\vectb{v}$ is the fluid velocity, $\mu\left(\dot{\gamma}\right)$ is the
  effective viscosity depending on the scalar measure $\dot{\gamma}$ of the strain-rate tensor
  \begin{equation}\label{eq:strain rate magn}
   \dot{\gamma}=\left|2\vectb{E}\ \right|=\sqrt{2\text{tr}\left( \vectb{E}^T\vectb{E} \right)}=\sqrt{2I_2}\ ,
  \end{equation}
  with $I_2$ the second principal trace of the infinitesimal strain-rate tensor
  \cite{irgens2014rheology, itskov2007tensor, bird1987dynamics}.

  \subsection{SRB model}
    \begin{figure}[!tb]
                \centering
                \includegraphics[width=0.98\textwidth]{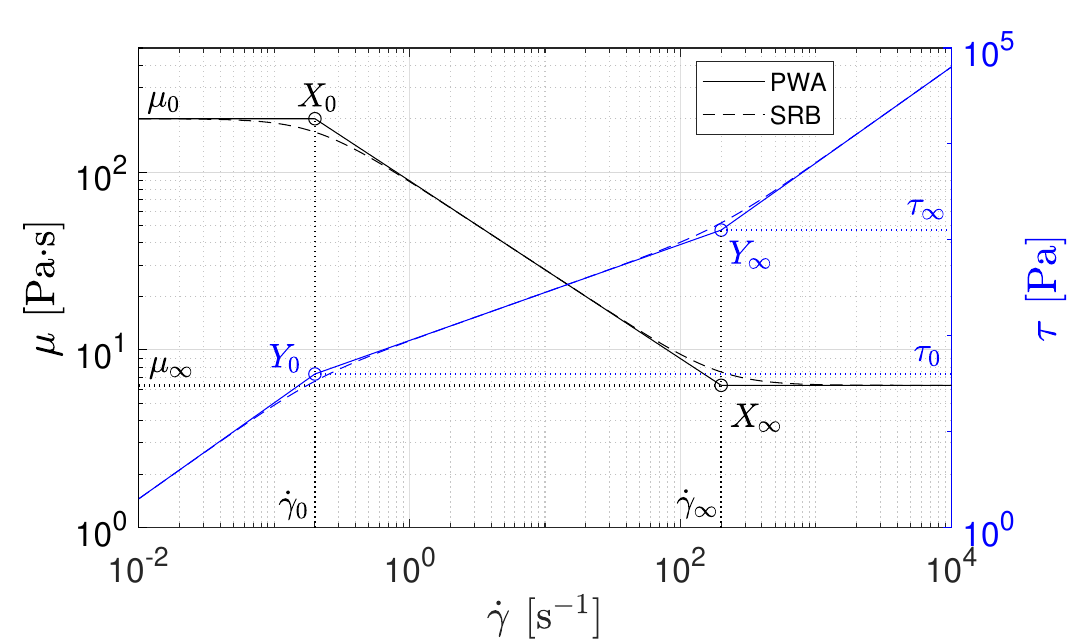}\hspace{+0.0cm}
                \caption{Examples of the viscosity response (black lines, left axis) and the
                corresponding deviatoric stress tensor norm (blue lines, right axis) predicted by
                the SRB model (dashed lines) and the PWA (continuous lines) as a function
                of the shear rate.
                Values of model parameters:
                $\mu_0 = 200$ Pa$\cdot$s, $\dot{\gamma}_0 = 1/\lambda_0 = 0.2$ s$^{-1}$,
                $\tau_0 = 40$ Pa, $\dot{\gamma}_\infty = 1/\lambda_\infty = 200$ s$^{-1}$,
                $n = 0.5$, $K = \mu_0\lambda_0^{n-1} = 89.4$ Pa$\cdot$s$^{0.5}$, $\mu_\infty = \mu_0 \left( \lambda_\infty/\lambda_0 \right)^{1-n} = 6.32$ Pa$\cdot$s,
                $\tau_\infty = 1264$ Pa, $a=2$.}
                \label{fig:SRB PWM mod examples}
    \end{figure}
  Due to the intrinsic issues of parameter identifiability of the Carreau-Yasuda
  model, which may lead to inaccurate physical interpretations and unreliable
  analytical flow solutions \cite{gallagher2019non, santesarti2025insight},
  the rheological response of ``Carreau-Yasuda-like'' fluids is described via \cite{santesarti2025insight}:
    \begin{equation}\label{eq:SRB model}
        \mu \left(\dot{\gamma}\right)=\mu_0\left[\frac{1+\left(\lambda_\infty\dot{\gamma}\right)^a}{1+\left(\lambda_0\dot{\gamma}\right)^a}\right]^\frac{\left(1-n\right)}{a}\ ,
    \end{equation} 
    where $\mu_0$ is the zero-shear rate viscosity (measured in [Pa$\cdot$s]),
    $n$ is the dimensionless power-law index such that $n\in (0,1)$, and $a$ is the
    dimensionless strictly-positive Yasuda parameter \cite{yasuda1979investigation}
    regulating the transition between the Newtonian and power-law regions.
    $\lambda_0$ and $\lambda_\infty$ are two time constants (measured in [s])
    delimiting the power-law region through the two characteristic shear rate
    levels ${\dot{\gamma}}_0=1/\lambda_0$ and ${\dot{\gamma}}_\infty=1/\lambda_\infty$,
    as shown in Fig. \ref{fig:SRB PWM mod examples}.
    For this reason, the model in Eq. \eqref{eq:SRB model} is referred to as
    the Shear Rate-Based model (SRB).
    The corresponding infinity-shear rate viscosity results $\mu_\infty=\mu_0\left( \lambda_\infty/\lambda_0 \right)^{1-n}$.
    This model can also describe an other possible rheological behaviour of
    shear-thinning fluids, corresponding to the Yasuda model \cite{yasuda1979investigation}.
    It applies to fluids with a viscosity characterized by an initial constant
    plateau at low shear rates followed by a shear-thinning branch at high shear rates.
    It is obtained from Eq. \eqref{eq:SRB model} in the limit for $\lambda_\infty\rightarrow 0^+$
    \begin{equation}\label{eq:SRB sub model 1}
        \mu\left(\dot{\gamma}\right)=\frac{\mu_0}{\left[1+\left(\lambda_0\dot{\gamma}\right)^a\right]^\frac{\left(1-n\right)}{a}}\ .
    \end{equation}
    
  \subsection{Power-law-based piecewise rheological approximation}\label{sec:rheol mods}
      The rheological response in Eq. \eqref{eq:SRB model} can be approximated with
      a power-law-based piecewise approximation (PWA) (see Fig. \ref{fig:SRB PWM mod examples})
      \begin{equation}\label{eq:PWM model visc}
            \mu \left(\dot{\gamma}\right) =
            \begin{cases}
              \mu_0               & \text{for $ \dot{\gamma} \le 1/\lambda_0 $} \\
              K\dot{\gamma}^{n-1} & \text{for $ 1/\lambda_0 < \dot{\gamma} \le 1/\lambda_\infty$} \\
              \mu_\infty          & \text{for $\dot{\gamma} > 1/\lambda_\infty$} \
            \end{cases}\ ,
      \end{equation}
      where $K=\mu_0\lambda_0^{n-1} = \mu_\infty\lambda_\infty^{n-1}$
      (measured in [Pa$\cdot$s$^n$]) is the consistency index.
      Using Eq. \eqref{eq:GNF definition} the deviatoric stress tensor norm
      results
      \begin{equation}\label{eq:PWM model stress}
            \tau \left(\dot{\gamma}\right) = 
            \mu\left(\dot{\gamma}\right)\dot{\gamma}=
            \begin{cases}
                  \mu_0 \dot{\gamma}      & \text{for $ \dot{\gamma} \le 1/\lambda_0, \tau \le \tau_0 $} \\
                  K\dot{\gamma}^{n}       & \text{for $ 1/\lambda_0 < \dot{\gamma} \le 1/\lambda_\infty, \tau_0 < \tau \le \tau_\infty$}\\
                  \mu_\infty \dot{\gamma} & \text{for $\dot{\gamma} > 1/\lambda_\infty, \tau > \tau_\infty$}
            \end{cases}\ ,
      \end{equation}
      where $\tau_0=\mu_0{\dot{\gamma}}_0$ and $ \tau_\infty=\mu_\infty{\dot{\gamma}}_{\infty\ }$ (measured in [Pa])
      are denoted as the zero-shear stress and the infinity-shear stress,
      respectively. 
      This approximation identifies three main viscosity regions depending on the 
      working shear rates and shear stresses applied to the fluid: an initial constant 
      Newtonian viscosity region characterized by $\mu_0$, an intermediate power-law
      viscosity region characterized by $K$ and $n$, and a final constant Newtonian 
      viscosity region characterized by $\mu_\infty$. 
      In the $\left( \mu,\, \dot{\gamma} \right)$ log-log graph, these three regions are
      represented by three lines intersecting at two points 
      \begin{equation}
          X_0=\left({\dot{\gamma}}_0,\mu_0\right),\ \ \ \ \ X_\infty=\left({\dot{\gamma}}_\infty,\mu_\infty\right),
      \end{equation}
      and in the $\left(\tau,\dot{\gamma}\right)$ log-log graph, at two points
      \begin{equation}
        Y_0=\left({\dot{\gamma}}_0,\tau_0\right),\ \ \ \ \ Y_\infty=\left({\dot{\gamma}}_\infty,\tau_\infty\right),
      \end{equation}
      as shown in Fig. \ref{fig:SRB PWM mod examples}.
  
\section{Mathematical modelling}\label{sec:semi-ana-mod-con} 
  \begin{figure}[!tb]
            \centering
            \includegraphics[width=0.8\textwidth]{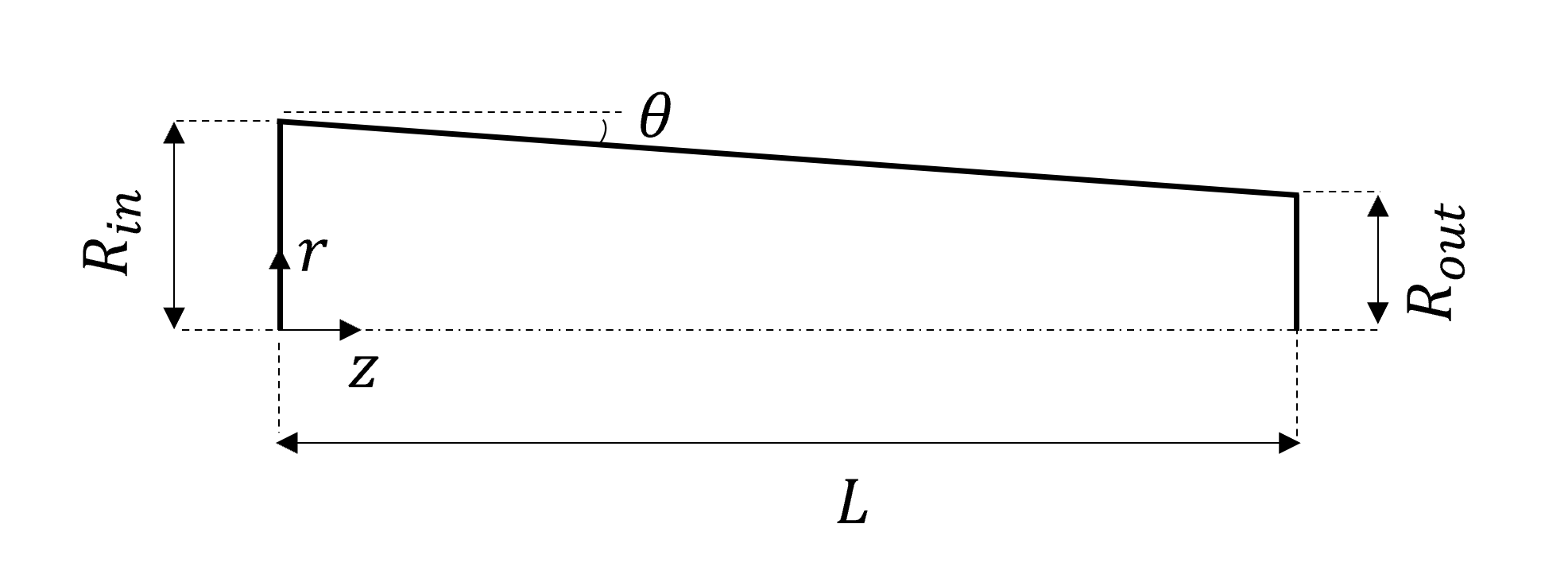}\hspace{+0.0cm}
            \caption{A tapered pipe geometry.}
            \label{fig:cone}
  \end{figure}
  By considering a flow in a slightly tapered pipe, the reference geometry is
  a conical duct (see Fig. \ref{fig:cone}) with a length $L$, an inlet and outlet
  radii $R_{in}$ and $R_{out}$, respectively, and a small opening angle $\theta$
  such that
  \begin{align}\label{eq:radius conical nozzle}
            &\tan \theta = \frac{R_{out}}{L} \left( \frac{1-\chi}{\chi} \right) = \theta +O(\theta^2) \simeq \theta, \\[10pt]
            &R(z) = R_{in} - z\tan \theta \simeq R_{in} - \theta z,
  \end{align}
  with $\chi = R_{out}/R_{in} \leq 1$.
  The cross section reduction implies an acceleration along the axis and a non-null
  radial velocity component to ensure the mass conservation.
  By using a cylindrical coordinate system and assuming an axial symmetric
  stationary flow, the velocity solution is $\vectb{v}=[v_r(r,z)\ 0\ v_z(r,z)]^T$.
  The mass and momentum conservation equations corresponding to a generic viscous
  inelastic fluid result:
  \begin{align}\label{eq:mass cons eq}
           \frac{1}{r}\derpar{}{r}(rv_r) + \derpar{v_z}{z} = 0\ ,
  \end{align}
  \begin{subequations}\label{eq:mom cons eq generic conical}
         \begin{align}
           r)      & \qquad  \rho \left( v_r \derpar{v_r}{r} + v_z\derpar{v_r}{z}  \right)  = \left[ \frac{1}{r}\derpar{}{r} \left( r\tau_{rr}\right) + \derpar{\tau_{zr}}{ z}-\derpar{\tau_{\theta  \theta}}{r}\right] - \derpar{p}{r}\ ,  \\[10pt]
           z)      & \qquad  \rho \left( v_r \derpar{v_z}{r} + v_z\derpar{v_z}{z}  \right)  = \left[ \frac{1}{r}\derpar{}{r} \left( r\tau_{zr}\right) + \derpar{\tau_{zz}}{ z}\right] - \derpar{p}{z} \ .
         \end{align}
  \end{subequations}
  
  \subsection{General relationships}
    Let us now consider the flow solution of a PWA fluid flowing within the channel.
    Since the PWA can be seen as a combination of a Newtonian and power-law model
    (see Section \ref{sec:rheol mods}) we aim at building a quasi-analytical solution
    by leaveraging on the well-known Newtoninan and power-law solutions based on
    the order-of-magnitude analysis, which are reported in Appendix A and B,
    respectively.
    In particular, the momentum balance equation (\ref{eq:mom cons eq generic conical}b)
    turns out (see Eqs. \eqref{eq:mom bal eqs con}, \eqref{eq:cap3QAM:mom bal eqs con powlaw})
    \begin{equation}\label{eq:mom bal eqs con inelastic}
          \frac{d p}{d z}  = \frac{1}{r}\derpar{\left( r\tau_{zr} \right)}{r} = 
          \frac{1}{r}\derpar{}{r} \left( r\mu \derpar{v_z}{r} \right)\ ,
    \end{equation}
    with a shear stress distribution corresponding to
    \begin{equation}\label{eq:shear stress con inelastic}
          \tau_{zr}(r,z) = \frac{d p}{dz}(z) \frac{r}{2}\ ,
    \end{equation}
    which is equivalent to the cylindrical flow case (see Appendix D), with the
    substantial difference of a non-constant and non-a priori known axial pressure
    gradient along the pipe axis.
    Furthermore, by integrating the Eq. \eqref{eq:shear stress con inelastic}
    the axial pressure gradient turns out
    \begin{equation}\label{eq:dpdz con inelastic}
          \frac{dp}{dz}(z) = \frac{4}{ R^2(z)} \int_0^{R(z)} \tau_{zr} (r,z) dr \ ,
    \end{equation}
    which shows how the axial pressure gradient value in each section
    of the duct is determined by the distribution of the shear stress
    along the section.

    Then, integrating Eq. \eqref{eq:mom bal eqs con inelastic} along the radius $R(z)$
    of a generic cross section, and applying the symmetric flow boundary
    condition  (i.e. $\partial v_z/\partial r |_{r=0} =0$) and the no-slip
    boundary condition at the wall (i.e. $v_z(r=R(z),z)=0$), the general solution
    of the axial velocity results
    \begin{equation}\label{eq:generalized ohm fluid law}
          v_z(r,z) = -\frac{dp}{dz}(z) \left[ F(r=R(z)) - F(r)  \right] \ ,
    \end{equation}
    with
    \begin{equation}\label{eq:hyd cond con flows F(r)}
          F(r) = \int \frac{r dr}{2\mu \left( \dot{\gamma}(r) \right)} \ ,
    \end{equation}
    wherein the dependence of the viscosity on the shear rate $\mu = \mu (\dot{\gamma})$ 
    has been highlighted.
    The corresponding flow rate turns out 
    \begin{equation}
        Q = \int_0^{R(z)} v_z 2\pi r dr 
        = -\frac{dp}{dz}(z) 2\pi \int_0^{R(z)} \left[ F(r=R(z)) - F(r)  \right] rdr\ .
    \end{equation}
    It is interesting nothing that the previous Eq. \eqref{eq:generalized ohm fluid law}
    for Generalized Newtonian fluids is analogous to the scalar Generalized
    Ohm's law
    \begin{equation}
          j=\sigma E \ ,
    \end{equation}
    where $j$ is the current density, $\sigma$ the electric conductivity and $E$
    the electric field.
    In particular, by comparing the two physical systems
    the axial pressure gradient $G=- dp/dz$ and the pressure drop $\Delta p$ 
    correspond to $E$ and the voltage drop $\Delta V$, respectively
    \begin{equation}
         \Delta p = \int_{0}^{L} G dz = \int_{0}^{L} - \frac{dp}{dz}dz \quad  \leftrightarrow \quad  |\Delta V| = \int_{0}^{L} E dx = \int_{0}^{L} -\frac{dV}{dx} dx\ .
    \end{equation}
    The velocity corresponds to the current density
    \begin{equation}
        v_z = \sigma_{hyd} G = \sigma_{hyd} \left( -\frac{dp}{dz} \right) \quad \leftrightarrow \quad 
        j = \sigma E \ ,
    \end{equation}
    where $\sigma_{hyd}$ is the hydraulic conductivity 
    \begin{equation}\label{eq:hyd cond con flows}
        \sigma_{hyd} = \left[ F(r=R(z)) - F(r)  \right] \ ,
    \end{equation}
    which depends on the fluid viscosity, the pipe geometry and also on the
    operating conditions, such as the flow rate, since it depends on
    the shear rate.

    Next, the general solution of the radial velocity derives from the mass
    conservation Eq. \eqref{eq:mass cons eq}
    \begin{equation}\label{eq:rad vel - mass cons eq}
      v_r(r,z) = - \frac{1}{r} \int r \derpar{v_z}{z}dr + \frac{f(z)}{r}\ .
    \end{equation}

  \subsection{Quasi-analytical solution}\label{sec:quasi-analytical sol}
    \begin{figure}[!tb]
      \centering
      \subfloat[][]{\includegraphics[width=.28\textwidth]{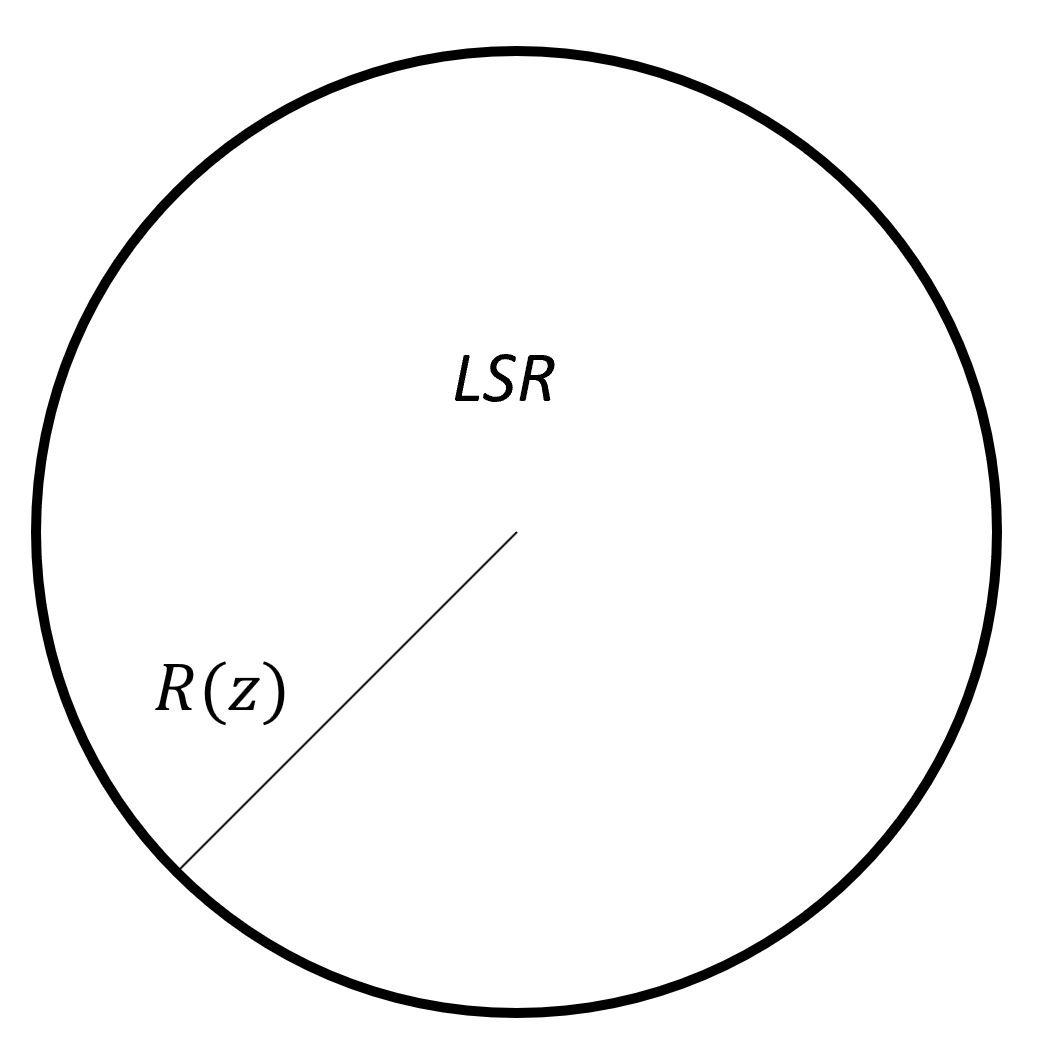}} \qquad
      \subfloat[][]{\includegraphics[width=.28\textwidth]{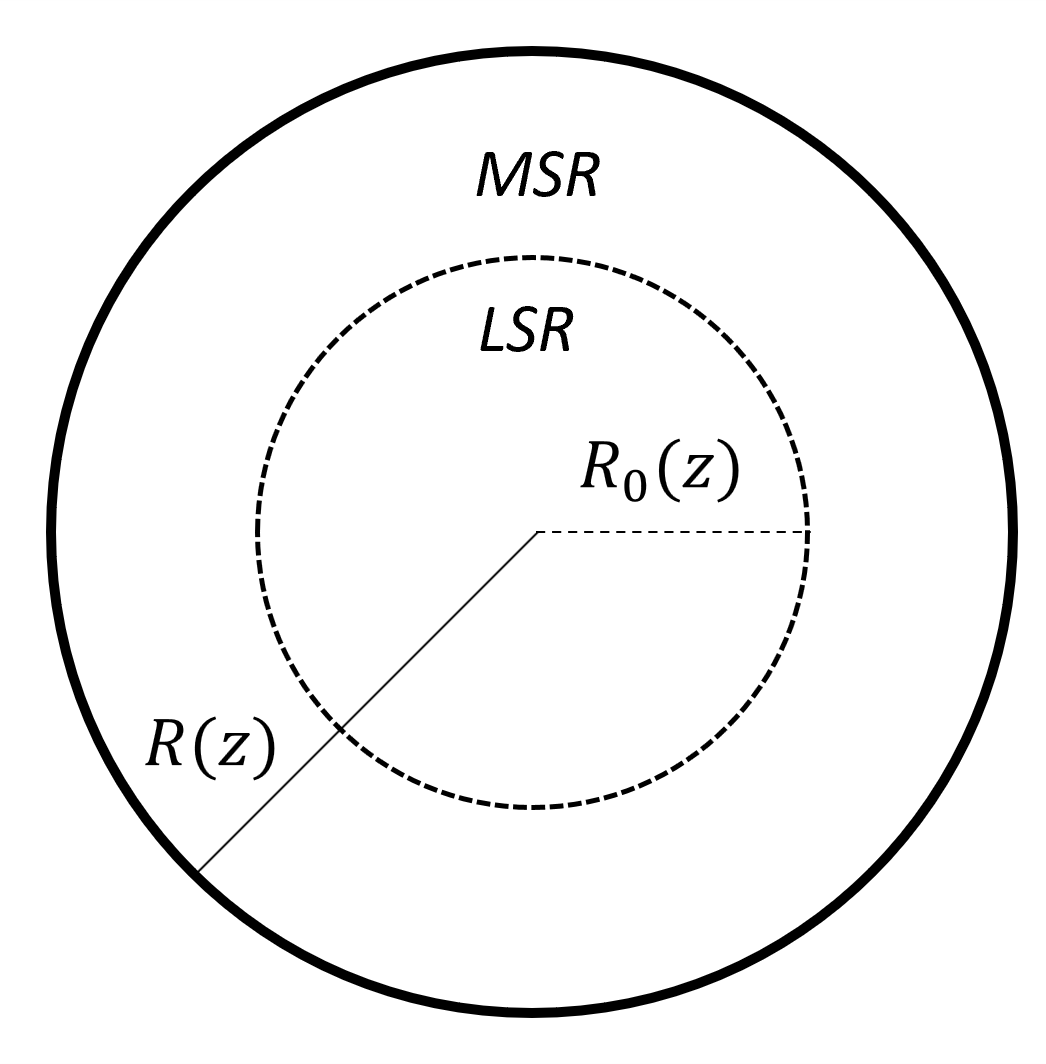}} \qquad
      \subfloat[][]{\includegraphics[width=.28\textwidth]{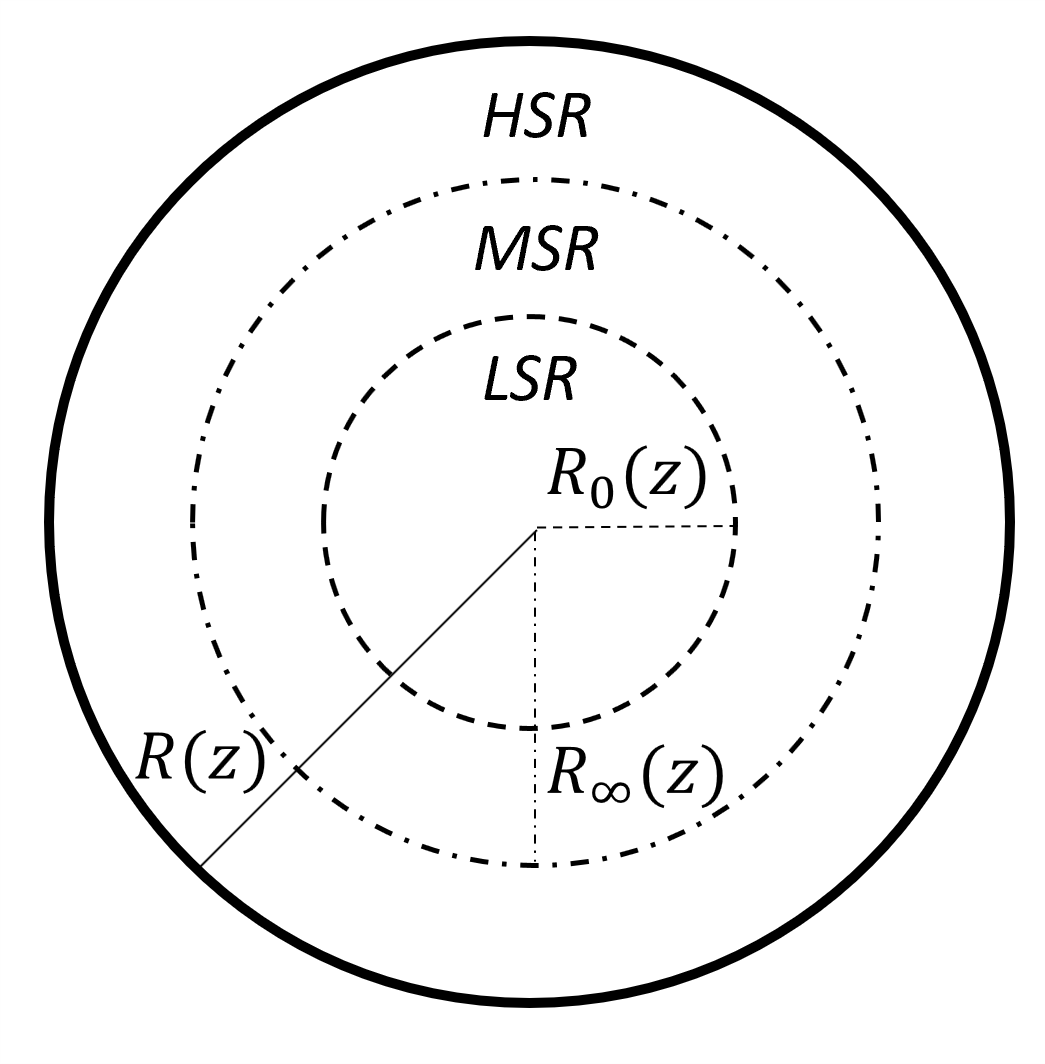}}
    	\caption{The three main flow conditions of PWA: (a) Low-shear rate flow,
    	(b) Medium-shear rate flow, (c) High-shear rate flow.}
     	\label{fig:LSR MSR HSR regimes conical}
    \end{figure}
    Given as known the pipe geometry (see Fig. \ref{fig:cone}), the imposed flow rate $Q$,
    and the rheological properties of the fluid, once the axial pressure gradient
    $p'(z)=dp(z)/dz$ has been evaluated through the iterative procedure presented in
    Section \ref{sec:iter proc}, from Eq. \eqref{eq:shear stress con inelastic}
    it is possible to determine the two radius functions $R_0(z)$ and $R_\infty(z)$
    delimiting the three annular sections of the three viscosity regions of PWA
    in Eq. \eqref{eq:PWM model visc} (see Fig. \ref{fig:LSR MSR HSR regimes conical}(c))
    \begin{equation}\label{eq:R0 and R infinity con}
      R_0(z)=\frac{2\tau_0}{\left|p'(z)\right|}\ ,\ \ \ \ \ \ \ \ \ \ R_\infty(z)=\frac{2\tau_\infty}{\left|p'(z)\right|}\ .
    \end{equation} 
    Compared to the cylindrical flow case (see Appendix D), they are non-constant
    along the streamwise direction, since the axial pressure gradient increases
    along the conical pipe axis.
    Depending on the operating conditions ($Q$, $\Delta p$), the pipe geometry
    ($R_{in}$, $R_{out}$, $L$, $\theta$), the rheological properties of the
    fluid ($\mu_0, \lambda_0, \lambda_\infty, n$), and on
    the axial coordinate $z$, three main flow conditions can develop as
    represented schematically in Fig. \ref{fig:LSR MSR HSR regimes conical}:
    
    \begin{itemize}
      \item
      a low-shear rate (LSR) flow where the fluid behaves entirely as a Newtonian
      fluid with viscosity $\mu_0$. The wall shear rate resuts
	  $\left|{\dot{\gamma}}_{wall}(z)\right|\le 1/\lambda_0$ and the wall
	  shear stress
	  $\left|\tau_{wall}(z)\right|\le\tau_0$ (Fig. \ref{fig:LSR MSR HSR regimes conical}(a));
	  \item
	  a medium-shear rate (MSR) flow where the fluid behaves as a Newtonian fluid
	  with viscosity $\mu_0$ for $r \le R_0(z)$, and as a power-law fluid with
	  constants $(K,n)$ for $R_0(z)< r \le R(z)$.
	  In this regime, it results with a
	  $1/\lambda_0<\left|{\dot{\gamma}}_{wall}(z)\right|\le 1/\lambda_\infty$
	  and a $\tau_0<\left|\tau_{wall}(z)\right|\le\tau_\infty$
	  (Fig. \ref{fig:LSR MSR HSR regimes conical}(b));
	  \item
	  and a high-shear rate (HSR) flow where the fluid behaves as a Newtonian
	  fluid with viscosity $\mu_0$ for $r \le R_0(z)$, as a power-law fluid with
	  constants $(K,n)$ for $R_0(z)< r \le R_\infty(z)$, and as a Newtonian fluid
	  with viscosity $\mu_\infty$ for $R_\infty (z) < r \le R(z)$.
	  The wall shear rate results $\left|{\dot{\gamma}}_{wall}(z)\right|> 1/\lambda_\infty$
	  and the wall shear stress $\left|\tau_{wall}(z)\right|>\tau_\infty$
	  (Fig. \ref{fig:LSR MSR HSR regimes conical}(c)).
    \end{itemize}
    For a given axial section, by increasing the pressure drop $\Delta p$,
    the $R_0(z)$ and $R_\infty(z)$ functions decrease, and the flow gradually
    passes from the condition in Fig. \ref{fig:LSR MSR HSR regimes conical}(a)
    to the one in Fig. \ref{fig:LSR MSR HSR regimes conical}(c).
    The expressions of the axial velocity, shear rate and flow rate derive from
    the integration of Eq. \eqref{eq:mom bal eqs con inelastic} along the radius.
    The expression of the radial velocity derives from the integration of the
    mass conservation in Eq. \eqref{eq:rad vel - mass cons eq}.
    The boundary conditions are the symmetric condition on the axis (i.e. $\partial v_z (0,z)/\partial r  =0$),
    the no-slip boundary condition at the wall (i.e. $v_z\left(R(z),z\right)=v_r\left(R(z),z\right)=0$),
    and the continuity conditions on the axial (i.e. $v_{z,LSR}\left(R_0(z),z\right)=v_{z,MSR}\left(R_0(z),z\right)$) and on the radial (i.e. $v_{r,LSR}\left(R_0(z),z\right) = v_{r,MSR} \left(R_0(z),z\right)$) velocity between each viscosity zone.
    The solutions of the problem for each flow regime read:
    \begin{itemize}
		\item 
		Low-shear rate (LSR) flow, $\left|\dot{\gamma}_{wall}(z)\right| \le 1/\lambda_0 ,\ \left|\tau_{wall}(z)\right|\le\tau_0$,
		\begin{fleqn}
        \begin{subequations}\label{eq:PWM-LSR expressions conical}
	    \begin{align}
		&\quad v_{z,LSR}\left(r,z\right) = -\frac{p'(z) R^2(z)}{4\mu_0}\left[1-\left(\frac{r}{R(z)}\right)^2\right],\\[10pt]
		&\quad v_{r,LSR}\left(r,z\right) = -v_{z,LSR}\left(r,z\right)\frac{\theta r}{R(z)},\\[10pt]
		&\quad \dot{\gamma}_{zr,LSR}\left(r,z\right) = \frac{p'(z)}{2\mu_0}r, \qquad Q=Q_{LSR} = -\frac{p'(z)\pi R^4(z)}{8\mu_0} ;
		\end{align}
        \end{subequations}
	    \end{fleqn}
        with a flow rate $Q = Q_{LSR}$.
		\item 
		Medium-shear rate (MSR) flow, $1/\lambda_0<\left|\dot{\gamma}_{wall}(z)\right|\le 1/\lambda_\infty, \tau_0<\left|\tau_{wall}(z)\right|\le\tau_\infty$,
      \begin{fleqn}
      \begin{subequations}\label{eq:PWM-MSR expressions conical}

        \text{for} $r \le R_0(z)$,
			\begin{align}
			&\quad v_{z,LSR}\left(r,z\right) = A_0(z)+\frac{p'(z)}{4\mu_0}r^2,  \\[10pt]
			&\quad v_{r,LSR}\left(r,z\right) = -\frac{r}{2}\left( A_0'(z) + \frac{p''(z)r^2}{8\mu_0} \right) , \\[10pt]
		    &\quad \dot{\gamma}_{zr,LSR}\left(r,z\right) = \frac{p'(z)}{2\mu_0}r, \qquad Q_{LSR} = \pi R_0^2(z)A_0(z)+\frac{p'(z)\pi R_0^4(z)}{8\mu_0} ;
			\end{align}

	    \text{for} $r > R_0(z)$,
	    	\begin{align}
	    	&\quad v_{z,MSR}\left(r,z\right) = \left(-\frac{p'(z)}{2K}\right)^\frac{1}{n} \frac{R(z)^\alpha-r^\alpha}{\alpha}, \\[10pt]
            &\quad v_{r,MSR}\left(r,z\right) = -r\left(-\frac{p'(z)}{2K}\right)^\frac{1}{n}\left[ \frac{ p''(z)}{p'(z)\alpha n} \left(  \frac{R^\alpha(z)}{2} - \frac{r^\alpha}{\alpha +2} \right) - \frac{R^{\alpha-1} (z) \theta }{2} \right] +\frac{f_0(z)}{r} ,  \\[10pt]
            &\quad \dot{\gamma}_{zr,MSR}\left(r,z\right)= -\left(-\frac{p'(z)}{2K}\right)^\frac{1}{n}r^{1/n}, \\[10pt]
            &\quad Q_{MSR}=\frac{2\pi}{\alpha}\left(-\frac{p'(z)}{2K}\right)^\frac{1}{n}\left[\frac{R^\alpha(z) \left(R^2(z) -R_0^2(z) \right)}{2}-\frac{R^\beta(z) - R_0^\beta(z) }{\beta}\right] ;
	    	\end{align}	
        \end{subequations}
	    \end{fleqn}
        with a flow rate $Q = Q_{LSR} + Q_{MSR}$.
		\item 
		High-shear rate (HSR) flow, $\left|\dot{\gamma}_{wall}(z)\right|> 1/\lambda_\infty , \ \left|\tau_{wall}(z)\right|>\tau_\infty$,
		
		for $r \le R_0(z)$, 
		\begin{fleqn}
		\begin{subequations}\label{eq:PWM-HSR expressions conical}
			\begin{align}
			&\quad v_{z,LSR}\left(r,z\right) = A_0(z) + \frac{p'(z)}{4\mu_0}r^2,  \\[10pt]
			&\quad v_{r,LSR}\left(r,z\right) = -\frac{r}{2}\left( A_0'(z) + \frac{p''(z)r^2}{8\mu_0} \right) , \\[10pt]
			&\quad \dot{\gamma}_{zr,LSR}\left(r,z\right) = \frac{p'(z)}{2\mu_0}r, \qquad Q_{LSR} = \pi R_0^2(z)A_0(z)+\frac{p'(z)\pi R_0^4(z)}{8\mu_0} ;
			\end{align}
			\text{for} $ R_0(z) < r \le R_\infty (z) $, 	    
			\begin{align}
			&\quad v_{z,MSR}\left(r,z\right)= -\frac{1}{\alpha}\left(-\frac{p'(z)}{2K}\right)^\frac{1}{n}r^\alpha+A_1(z), \\[10pt]
			&\quad v_{r,MSR}\left(r,z\right) = \frac{1}{n\alpha\beta} \left(-\frac{p'(z)}{2K}\right)^\frac{1}{n} \frac{ p''(z)}{p'(z)} r^{\alpha+1} - \frac{A_1'(z)r}{2} + \frac{f_0(z)}{r},  \\[10pt]
			&\quad \dot{\gamma}_{zr,MSR}\left(r,z\right)=  -\left(-\frac{p'(z) }{2K}\right)^\frac{1}{n}r^\frac{1}{n}, \\[10pt]
			&\quad  Q_{MSR}=A_1(z)\pi\left(R_\infty^2(z)-R_0^2(z)\right)-\frac{2\pi}{\alpha\beta}\left(-\frac{p'(z)}{2K}\right)^\frac{1}{n}\left(R_\infty^\beta(z)-R_0^\beta (z)\right) ;
			\end{align}	
			\text{for} $ r > R_\infty (z)$, 	    
			\begin{align}
			&\quad v_{z,HSR}\left(r,z\right)= -\frac{p'(z)}{4\mu_\infty}\left(R^2(z)-r^2\right), \\[10pt]
			&\quad v_{r,HSR}\left(r,z\right)= \frac{p''(z)}{16\mu_\infty}\left(2R^2(z)-r^2\right)r - \frac{\theta p'(z)R(z) }{4\mu_\infty}r + \frac{f_1(z)}{r},   \\[10pt]
			&\quad \dot{\gamma}_{zr,HSR}\left(r,z\right)= \frac{p'(z)}{2\mu_\infty}r, \qquad Q_{HSR}=-\frac{p'(z)\pi}{8\mu_\infty}\left(R^2(z)-R_\infty^2(z)\right)^2 ;
			\end{align}
        \end{subequations}
		\end{fleqn}
		with a flow rate $Q = Q_{LSR} + Q_{MSR} + Q_{HSR}$.
	
    \end{itemize}
    where $p'(z)= dp(z)/dz$, $\alpha=(n+1)/n$, and $\beta=(3n+1)/n$.
    The integration functions of the axial velocity $A_0(z), A_0'(z), A_1(z)$
    and $ A_1'(z)$, and of the radial velocity $f_0(z)$ and $f_1(z)$,
    derive by the continuity conditions between each viscosity zone and are
    reported in Appendix C.
    The two radius functions $R_0 (z)$ and $R_\infty (z)$ are not known, but they
    can be evaluated from the Eqs. \eqref{eq:R0 and R infinity con} once $p'(z)$
    has been determined (see Section \ref{sec:iter proc}).
    Differently from the conical flow cases of a Newtonian and a power-law fluid
    (see Appendix A and Appendix B, respectively), it is not possible to obtain
    a closed-form relationship between the flow rate $Q$ and the axial pressure
    gradient $p'(z)$.
    Therefore, an iterative semi-analytical procedure is needed.

    \subsection{Iterative procedure for determining the pressure gradient distribution $p'(z)$}\label{sec:iter proc}
          \begin{figure}[]
            \centering
            \includegraphics[width=0.8\textwidth]{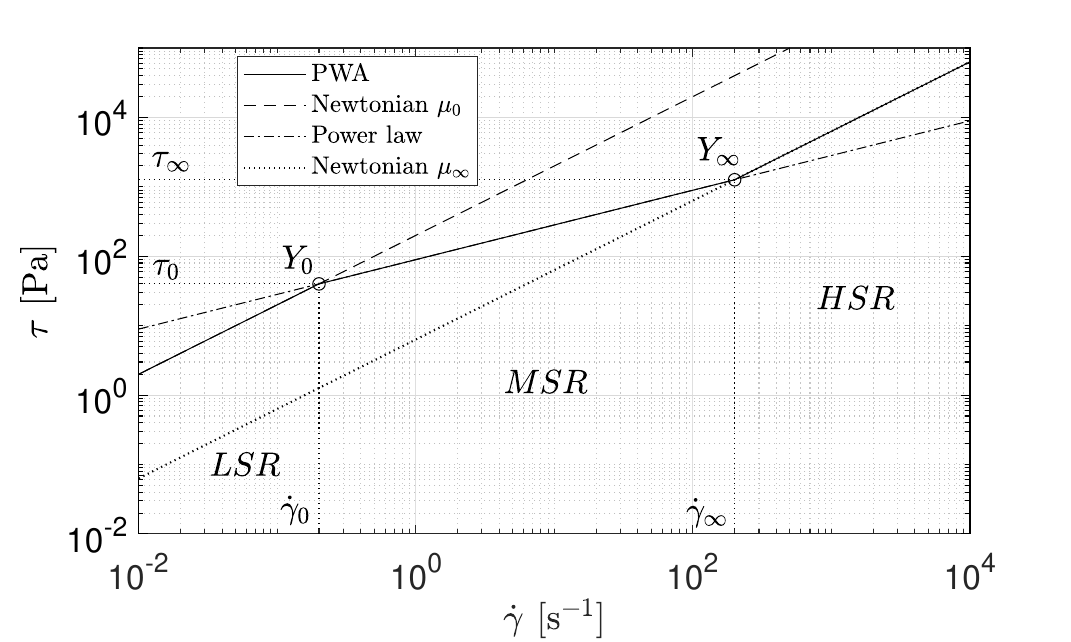}\hspace{+0.0cm}
            \caption{
            Example of the PWA shear stress distribution (continuous line) as a
            function of the shear rate to compute $p'(z)$ through the Algorithm \ref{alg:dpdz PWM}.
            The corresponding Newtonian and power-law sub-cases (dashed and dotted lines)
            with the three main viscosity regions (LSR, MSR and HSR) are highlighted.
            Values of model parameters:
            $\mu_0 = 200$ Pa$\cdot$s, $\dot{\gamma}_0 = 1/\lambda_0 = 0.2$ s$^{-1}$,
            $\tau_0 = 40$ Pa, $\dot{\gamma}_\infty = 1/\lambda_\infty = 200$ s$^{-1}$, $n = 0.5$,
            $K = 89.4$ Pa$\cdot$s$^{0.5}$, $\mu_\infty = 6.32$ Pa$\cdot$s,
            $\tau_\infty = 1264$ Pa.
            }
            \label{fig:PWM newto-powlaw example}
          \end{figure}

          The following procedure implemented in the MATLAB environment (R2024b, MathWorks, MA, USA)
          computes the axial pressure gradient $p'(z)$
          of a PWA fluid given the pipe geometry, the flow rate $Q$,
          and the rheological properties of the fluid. Importantly, the axial pressure
          gradients $p'(z)|_{\mu_0}$, $p'(z)|_{\mu_\infty}$ (see Eq. \eqref{eq:flow rate eq con newto}
          with $\mu_0$ and $\mu_\infty$, respectively) and $p'(z)|_{(K,n)}$
          (see Eq. \eqref{eq:flow rate eq con powlaw}) of the corresponding
          Newtonian and power-law sub-cases of the PWA are determined and used
          as basis solutions.
          The flow regime of the PWA (LSR, MSR or HSR) in each axial section $z$
          of the nozzle is determined by the wall shear stress $\left|\tau_{wall}(z)\right|$
          value whether it is smaller or larger than the $\tau_0$ and
          $\tau_\infty$ values of the fluid (see Fig. \ref{fig:PWM newto-powlaw example}).
          Next, by considering the Eq. \eqref{eq:dpdz con inelastic}
          and looking to the stress plot example in Fig. \ref{fig:PWM newto-powlaw example},
          whatever the operating wall shear stress $ \tau_{wall}(z)=\tau_{max}(z)$,
          the shear stress distribution $\tau_{zr}(r,z)$ of the PWA along the
          cross section, and consequently the axial pressure gradient value, is lower
          or at most equal to the corresponding sub-cases of the Newtonian fluid
          with $\mu_0$ viscosity and power-law fluid with $(K,n)$ constants
          \begin{equation}\label{eq:tau PWM relation Newto-PL}
            \left| p'(z)\right|_{\text{PWA}} \le \left\{ \left| p'(z) \right|_{\mu_0}, \left| p'(z)\right|_{(K,n)} \right\}\ .
          \end{equation}
          Algorithm \ref{alg:dpdz PWM} computes the axial pressure gradient
          $p'(z)_{\text{PWA}}$ by using the conservation of the flow rate $Q$
          in each axial section, the solutions of the corresponding Newtonian
          and power-law sub-cases, and the previous inequality \eqref{eq:tau PWM relation Newto-PL}.
          Here, $\left| p'(z)\right|_{0}$ is the first guess value of the
          iterative root-finding equation
          \begin{equation}\label{eq:root finding equation}
            p'(z) \quad \text{such that} \quad Q_{LSR}\left[ p'(z) \right] + Q_{MSR}\left[ p'(z) \right] + Q_{HSR}\left[ p'(z) \right] - Q = 0\ .
          \end{equation}
          At the inlet, the minimum axial pressure gradient value between the
          Newtonian sub-case with viscosity $\mu_0$ , and power-law sub-case with $(K,n)$
          is first evaluated. Then, it is assed whether the corresponding operating
          wall shear stress $|\tau_{wall}|_{in}=|p'(z_{in})|_{\text{min}} R_{in}/2$
          is between or above the shear stress values of the two PWA characteristic
          points $Y_0$ and $Y_\infty$ (see Fig. \ref{fig:PWM newto-powlaw example}),
          and in this case the $|p'(z)_{\text{PWA}}|$ value is found through the Eq. \eqref{eq:root finding equation}.
          Next, the $|p'(z_i)_{\text{PWA}}|$ values along the pipe axis are
          similarly evaluated, by taking as first guess value that found in
          the previous cross section at $z_{i-1}$, owing to the increasing
          monotonicity of $|p'(z)|$.

%
%
           \RestyleAlgo{ruled}
            \SetKwComment{Comment}{/* }{ */}
            \begin{algorithm}[]
            \caption{Algorithm to compute the axial pressure gradient of the
            quasi-analytical solution in Eqs. \eqref{eq:PWM-LSR expressions conical}-\eqref{eq:PWM-HSR expressions conical}.}\label{alg:dpdz PWM}
            \vspace{4pt}
            The pipe axis $z$ is divided in $n$ intervals such that
            $\Delta z = z_{i+1}-z_{i}$, with $i=1,2,...,n+1$, $z_1=z_{in}$ and $ z_{n+1}=z_{out}$.\\
            \vspace{4pt}

            \textbf{At the inlet:} $z_i=z_1=z_{in}$\\
            \vspace{4pt}
            $\left| p'(z_{in})\right|_{min}=\text{min} \left\{ \left| p'(z_{in}) \right|_{\mu_0}, \left| p'(z_{in})\right|_{(K,n)} \right\} $ \\
            \vspace{4pt}
            {\uIf{  $ \left| p'(z_{in})\right|_{min} \le \frac{2\tau_0}{R_{in}}$ }{
               \vspace{6pt}
               $\left| p'(z_{in})\right|_{\text{PWA}} = \left| p'(z_{in}) \right|_{\mu_0} $ \\
               \vspace{4pt}
            }

            {\uElseIf{ $ \frac{2\tau_0}{R_{in}} < \left| p'(z_{in})\right|_{min} \le \frac{2\tau_\infty}{R_{in}}$ }{
               \vspace{6pt}
               $\left| p'(z_{in})\right|_{\text{PWA}} \quad : \quad Q_{LSR}(z_{in}) + Q_{MSR}(z_{in}) - Q = 0$ \\
               \vspace{4pt}
               with $\left| p'(z_{in})\right|_0=\left| p'(z_{in})\right|_{min} $
               }
            }

            \vspace{4pt}
            {\ElseIf{  $\left| p'(z_{in})\right|_{min} > \frac{2\tau_\infty}{R_{in}}$ }{
               \vspace{6pt}
               $\left| p'(z_{in})\right|_{\text{PWA}} \quad : \quad Q_{LSR}(z_{in}) + Q_{MSR}(z_{in}) + Q_{HSR}(z_{in}) - Q = 0$ \\
               \vspace{4pt}
               with $\left| p'(z_{in})\right|_0=\left| p'(z_{in})\right|_{\mu_\infty} $
               }
            }
            }
            (continuing)
            \end{algorithm}

            \begin{algorithm}[]
            \ContinuedFloat
            \caption{Algorithm to compute the axial pressure gradient of the
            quasi-analytical solution in Eqs. \eqref{eq:PWM-LSR expressions conical}-\eqref{eq:PWM-HSR expressions conical} (continued).}
            \vspace{4pt}

            \textbf{Along the axis:} $ z_{in} < z_i \le z_{out}$\\
            \uIf{$\left| p'(z_{i-1})\right|_{\mathrm{PWA}} \le \frac{2\tau_0}{ R(z_{i-1}) }$ }{
                \vspace{6pt}
               $\left| p'(z_{i})\right|_{\text{PWA}} = \left| p'(z_{i}) \right|_{\mu_0} $ \\
               \vspace{4pt}
               \If{$\left| p'(z_{i})\right|_{\mathrm{PWA}} > \frac{2\tau_0}{ R(z_{i}) }$}{
               \vspace{6pt}
               $\left| p'(z_{i})\right|_{\text{PWA}} \quad : \quad Q_{LSR}(z_{i}) + Q_{MSR}(z_{i}) - Q = 0$ \\
               \vspace{4pt}
               with $\left| p'(z_{i})\right|_0=\left| p'(z_{i-1})\right|_{\text{PWA}} $
               }
            }

            {\uElseIf{$ \frac{2\tau_0}{ R(z_{i-1})} < \left| p'(z_{i-1})\right|_{\mathrm{PWA}} \le \frac{2\tau_\infty}{ R(z_{i-1}) }$}{
                \vspace{6pt}
                $\left| p'(z_{i})\right|_{\text{PWA}} \quad : \quad Q_{LSR}(z_{i}) + Q_{MSR}(z_{i}) - Q = 0$ \\
                \vspace{4pt}
                with $\left| p'(z_{i})\right|_0=\left| p'(z_{i-1})\right|_{\text{PWA}} $\\
                \vspace{4pt}

                \If{$\left| p'(z_{i})\right|_{\mathrm{PWA}}>\frac{2\tau_\infty}{ R(z_{i}) }$}{
                \vspace{6pt}
               $\left| p'(z_{i})\right|_{\text{PWA}} \quad : \quad Q_{LSR}(z_{i}) + Q_{MSR}(z_{i}) + Q_{HSR}(z_{i}) - Q = 0$ \\
               \vspace{4pt}
               with $\left| p'(z_{i})\right|_0=\left| p'(z_{i-1})\right|_{\text{PWA}} $
                }
              }
            }

            {\ElseIf{$ \left| p'(z_{i-1})\right|_{\mathrm{PWA}} > \frac{2\tau_\infty}{ R(z_{i-1}) }$}{
                \vspace{6pt}
               $\left| p'(z_{i})\right|_{\text{PWA}} \quad : \quad Q_{LSR}(z_{i}) + Q_{MSR}(z_{i}) + Q_{HSR}(z_{i}) - Q = 0$ \\
               \vspace{4pt}
               with $\left| p'(z_{i})\right|_0=\left| p'(z_{i-1})\right|_{\text{PWA}} $
                }
            }
            \end{algorithm}

\section{Results and Discussion}\label{sec:results}      
  The quasi-analytical solution in Eqs. \eqref{eq:PWM-LSR expressions conical}-\eqref{eq:PWM-HSR expressions conical}
  has been applied for the analysis of extrusion bioprinting
   \cite{conti2022models, ozbolat2016current}.
  The reference geometry is a conical pipe with an inlet and outlet radii 
  $R_{in}=1.5$ mm and $R_{out}=0.25$ mm, respectively, a length $L=20$ mm,
  and a corresponding taper angle $\theta = 3.58^{\circ}$ (see Fig. \ref{fig:cone}). 
  The mean extrusion velocity $\bar{V}_{out}$ at the outlet section ranges between $0-40$ mm/s.
  
  \subsection{Numerical verification}\label{sec:results:num ver}
    The quasi-analytical solution has been verified by simulating the extrusion 
    process through numerical solutions. The CFD simulations have been performed in
    Ansys Fluent by reproducing the axisymmetric pipe geometry in Fig. \ref{fig:cone}
    and solving the governing Eqs. \eqref{eq:mass cons eq}-\eqref{eq:mom cons eq generic conical}.
    A pressure-based coupled solver based on a fully implicit method for pressure
    gradient terms and face mass fluxes has been adopted.
    As regards the spatial discretization technique, the least squares cell based
    method has been employed for the computation of the spatial gradients, and a
    second order upwind scheme has been adopted for computing convection terms
    at cell faces.
    The boundary conditions applied read: power-law velocity profiles at the
    inlet section for the axial and radial components according to Eqs. \eqref{eq:ax vel powlaw}, \eqref{eq:rad vel powlaw};
    a pressure-outlet condition with a $p_{out} = 0$ value at the outlet section;
    a no-slip condition at the wall; and an axisymmetric velocity condition at
    the pipe axis.

\begin{figure}[!tb]
    	\centering
    	\subfloat[viscosity]{\includegraphics[width=.46\textwidth]{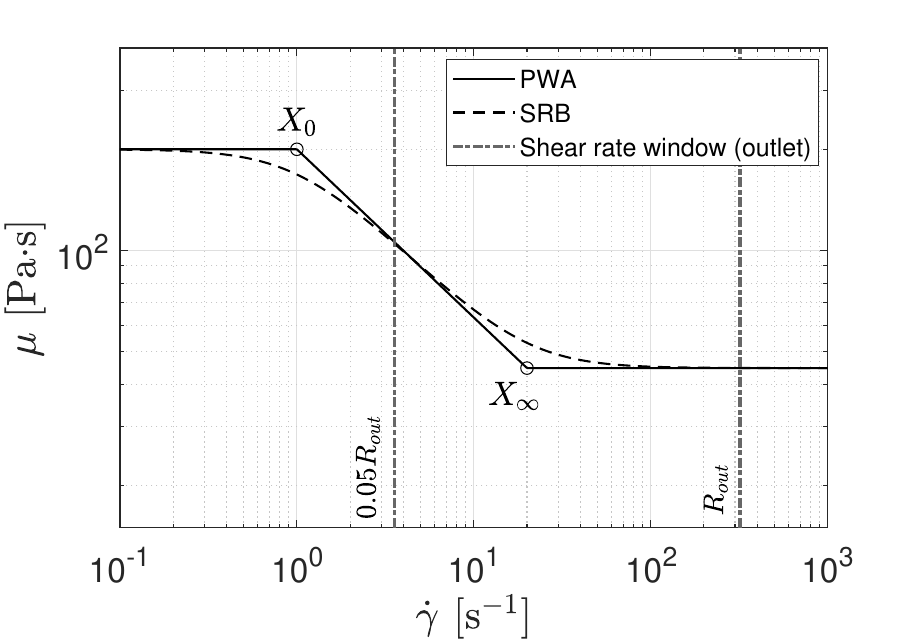}} \qquad
    	\subfloat[axis - percentage radii of viscosity annular sections (in Eqs. \eqref{eq:R0 and R infinity con})]{\includegraphics[width=.46\textwidth]{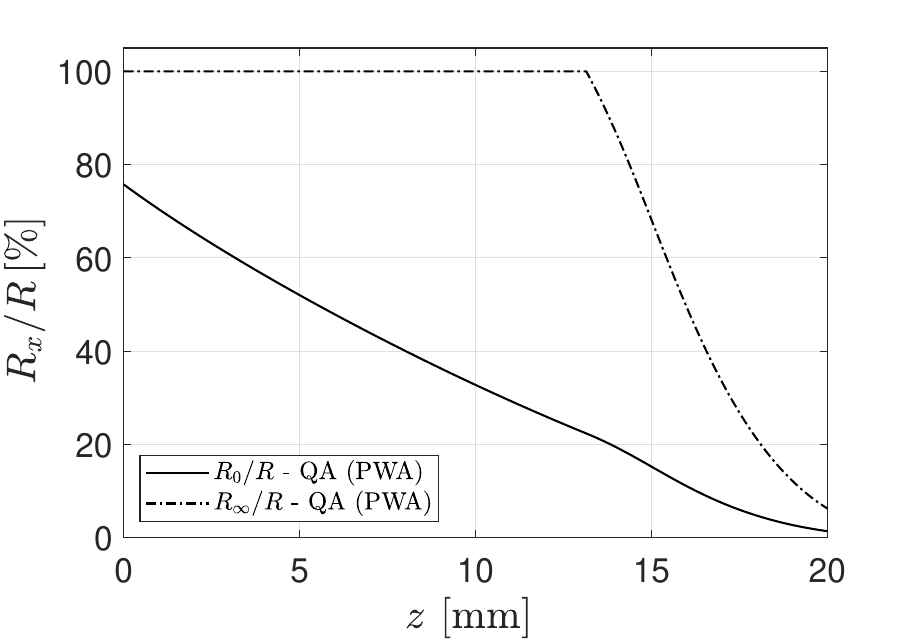}} \\
    	\subfloat[axis - pressure]{\includegraphics[width=.46\textwidth]{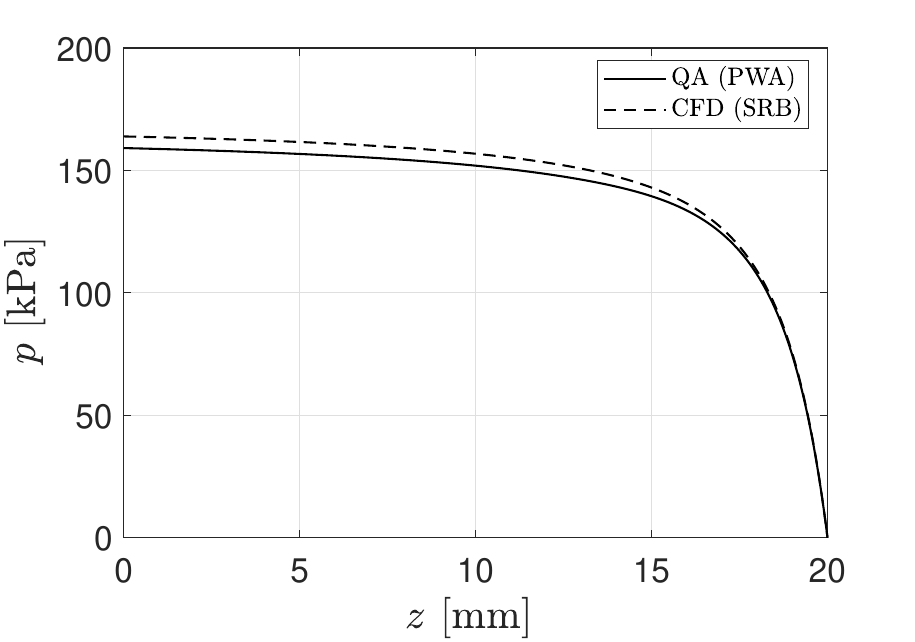}} \qquad
    	\subfloat[axis - axial velocity]{\includegraphics[width=.46\textwidth]{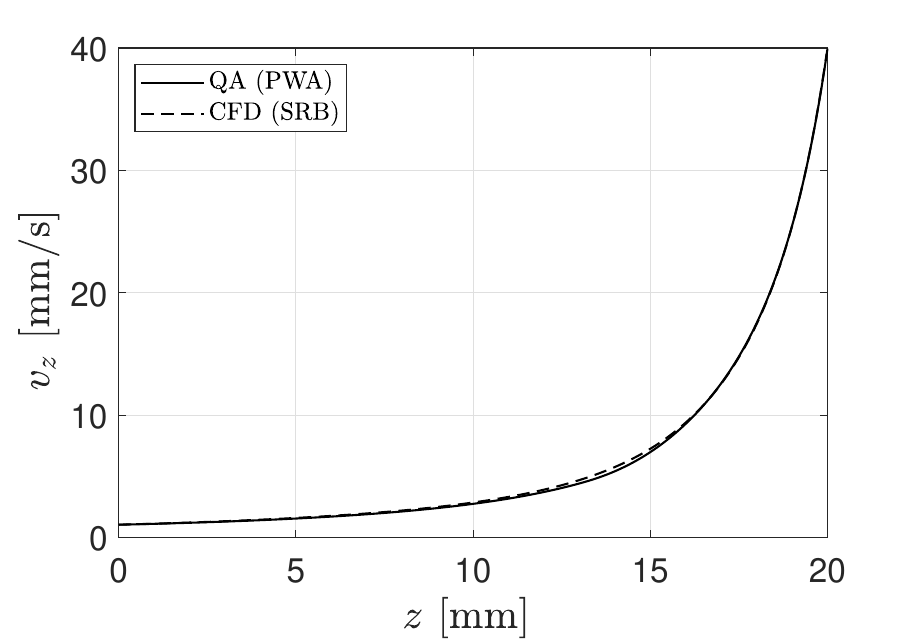}} \\
    	\subfloat[outlet - axial velocity]{\includegraphics[width=.46\textwidth]{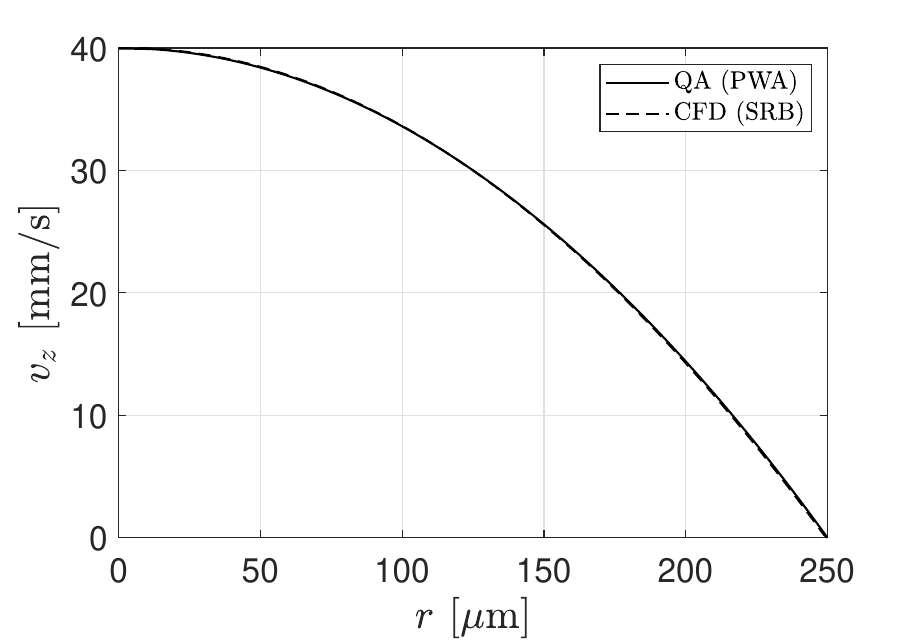}} \qquad
    	\subfloat[outlet - radial velocity]{\includegraphics[width=.46\textwidth]{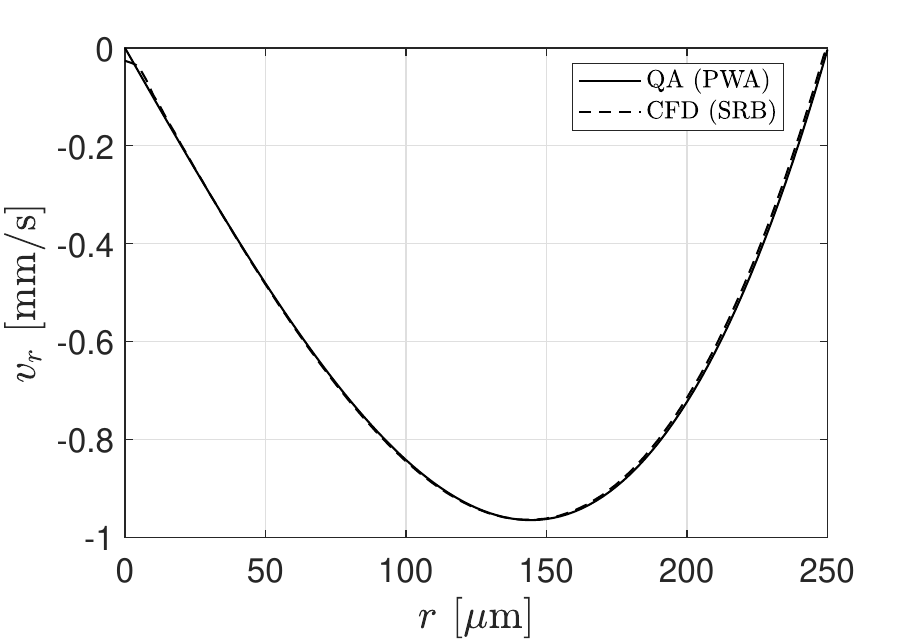}} \\
    	\caption{Numerical verification of the quasi-analytical solution (QA (PWA))
    	with comparison to the numerical solution (CFD (SRB)) applied to the
    	fluid A in Table \ref{tab:fluid rheo param values}.}
    	\label{fig:num ver fluid A}
    \end{figure}

\begin{figure}[!tb]
    	\centering
    	\subfloat[viscosity]{\includegraphics[width=.46\textwidth]{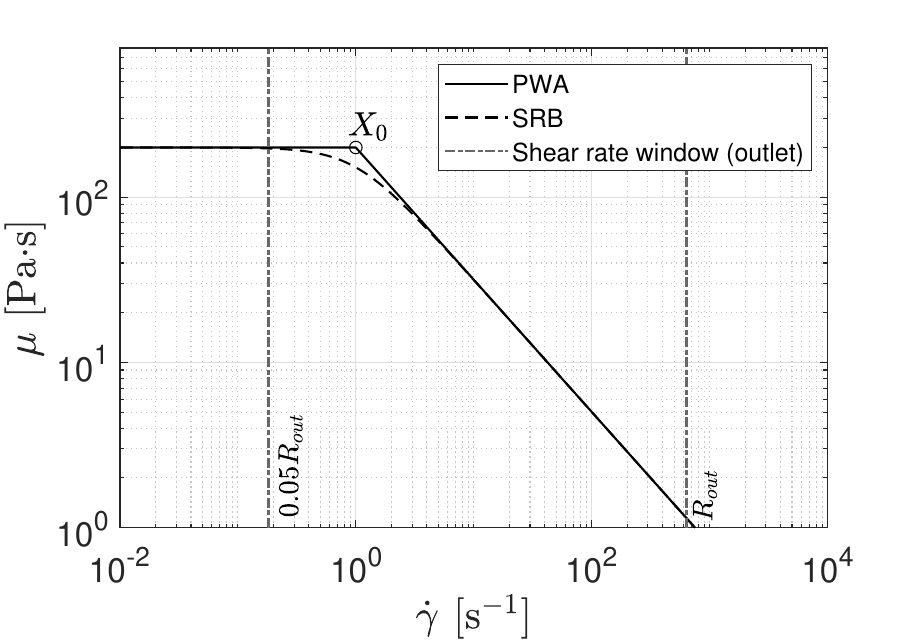}} \qquad
    	\subfloat[axis - percentage radius of Newtonian annular section (in Eq. \eqref{eq:R0 and R infinity con})]{\includegraphics[width=.46\textwidth]{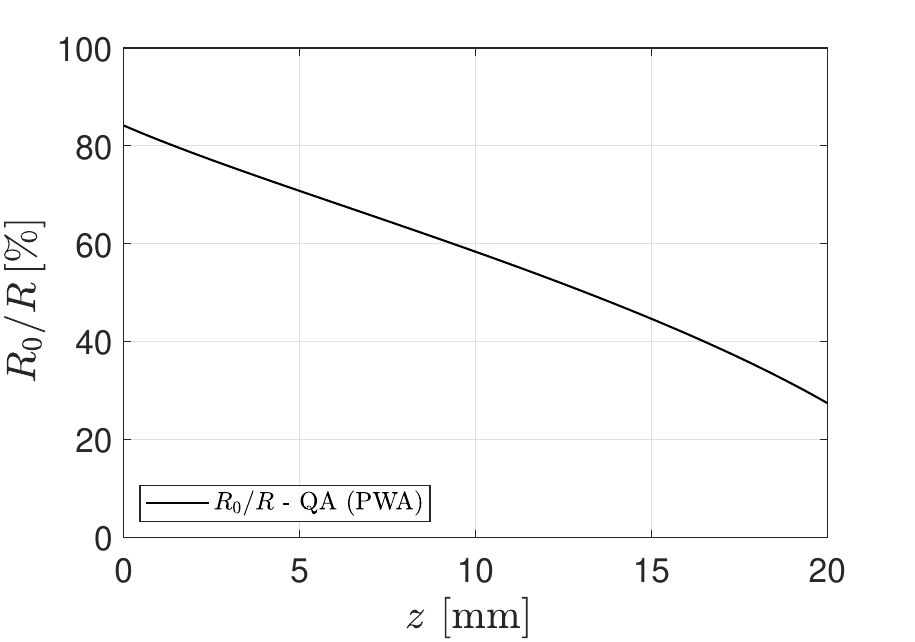}} \\
    	\subfloat[axis - pressure]{\includegraphics[width=.46\textwidth]{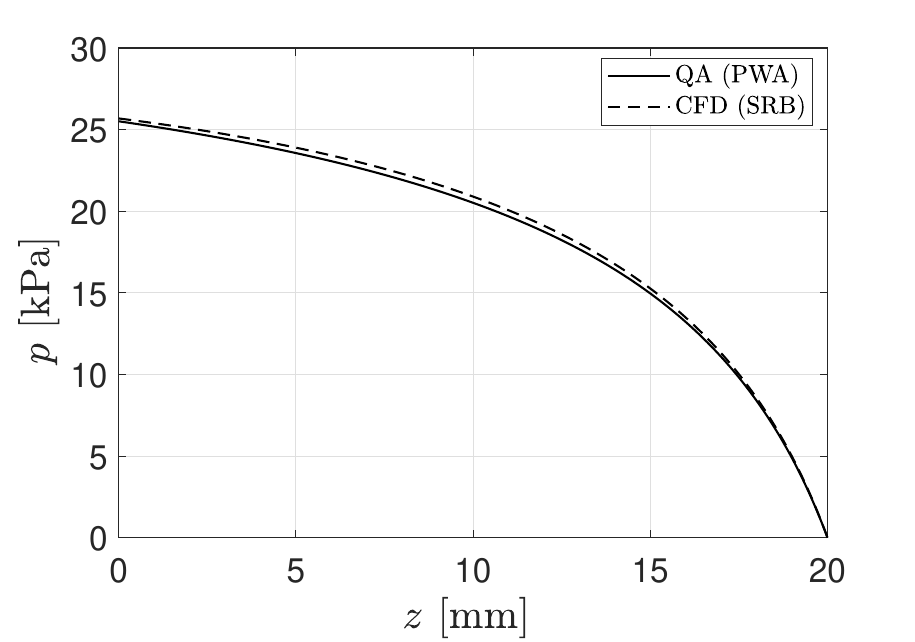}} \qquad
    	\subfloat[axis - axial velocity]{\includegraphics[width=.46\textwidth]{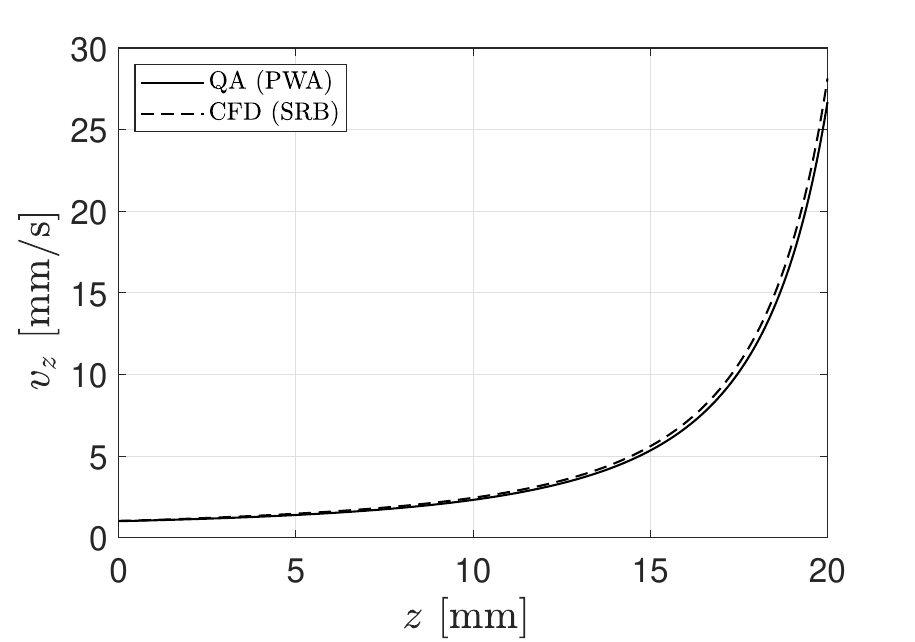}} \\
    	\subfloat[outlet - axial velocity]{\includegraphics[width=.46\textwidth]{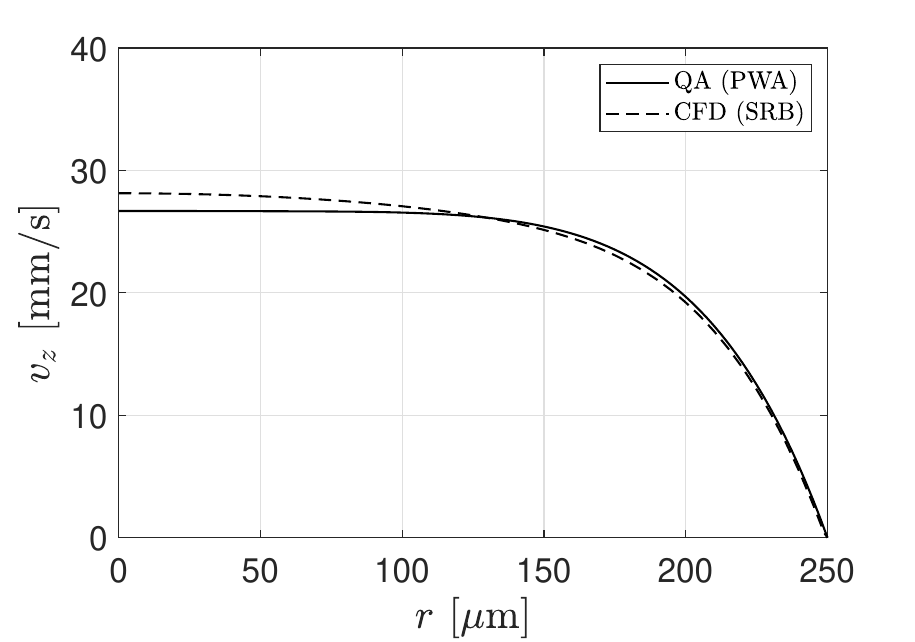}} \qquad
    	\subfloat[outlet - radial velocity]{\includegraphics[width=.46\textwidth]{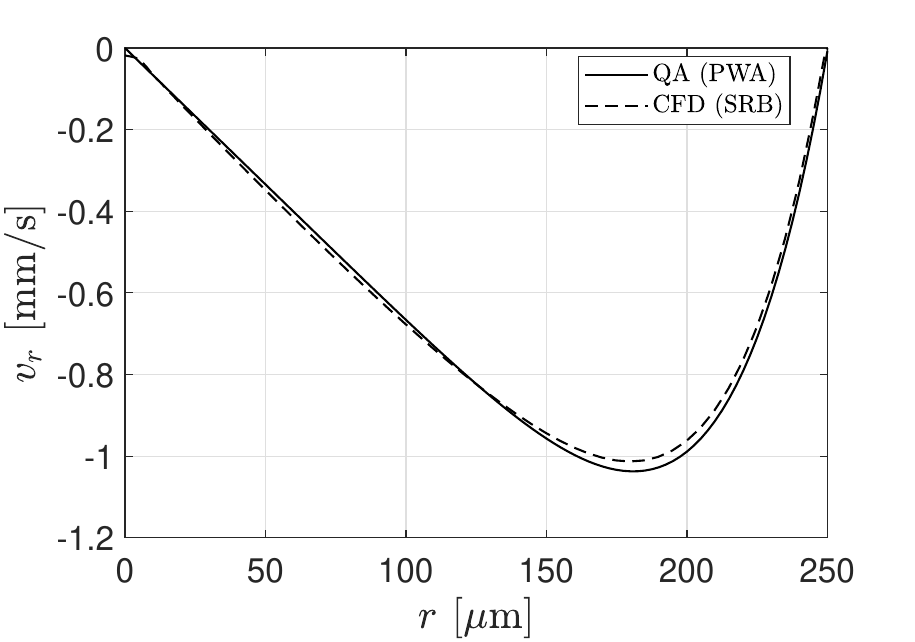}} \\
    	\caption{Numerical verification of the quasi-analytical solution (QA (PWA))
        with comparison to the numerical solution (CFD (SRB)) applied to the fluid
        B in Table \ref{tab:fluid rheo param values}.}
    	\label{fig:num ver fluid B}
    \end{figure}

    Two non-Newtonian inelastic representative fluids with rheological properties 
    similar to alginate-based bio-inks \cite{paxton2017proposal, kiyotake2019development} 
    and described through the SRB model in Eq. \eqref{eq:SRB model} and the 
    corresponding case of Yasuda model in Eq. \eqref{eq:SRB sub model 1} have been considered
    (see Figs. \ref{fig:num ver fluid A}(a) and \ref{fig:num ver fluid B}(a)). 
    The values of rheological parameters are reported in Table \ref{tab:fluid rheo param values}.
    \begin{table}[!htb]
          \caption{Values of rheological parameters of the SRB model ($a=2$) and
          PWA of the two representative fluids analysed in Figs. \ref{fig:num ver fluid A}(a) 
          and \ref{fig:num ver fluid B}(a).}
          \centering 
          \vspace{5pt}
          \begin{tabular}{ccc ccc c }
	  		\hline
                     & $\mu_{0}$                & $\mu_\infty$              &  $\lambda_{0}$ & $\lambda_{\infty}$ & $n$          & $K$               \\ [0.5ex]
                     &  $[\text{Pa $\cdot$ s}]$ &  $[\text{Pa $\cdot$ s}]$  & $[\text{s}]$   & $[\text{s}]$       & $[-]$        & [Pa$\cdot$s$^{n}$]  \\
	  		\hline \vspace{-12pt}  \\
            fluid A  & $200$  &  $44.7$ & $1$ & 0.05 & 0.5 & $200$ \\[0ex]
	  		\hline
	  		fluid B  & $200$  &  -      & $1$ &   -  & 0.2 & $200$ \\[0ex]
	  		\hline
         \end{tabular}
         \label{tab:fluid rheo param values}
    \end{table}
    For both cases, a flow rate $Q=3.93$ mm$^3$/s corresponding to a mean extrusion
    velocity $\bar{V}_{out}=20$~mm/s has been applied.
    The axial pressure gradient $p'(z)$ value along the pipe axis obtained
    from Algorithm \ref{alg:dpdz PWM} is employed to reconstruct the pressure field $p(z)$
    and velocity solutions $v_r (r,z)$ and $v_z(r,z)$, and compared with numerical results.
    As shown in Figs. \ref{fig:num ver fluid A}(b) and \ref{fig:num ver fluid B}(b), 
    both cases analysed in Table \ref{tab:fluid rheo param values} are characterized 
    by a mixed Newtonian-power-law flow along the nozzle axis, with varying percentage
    radii of viscosity annular sections determined from Eqs. \eqref{eq:R0 and R infinity con}.
    The comparison of the pressure field along the pipe axis between the quasi-analytical 
    solution based on the PWA in Eq. \eqref{eq:PWM model visc} (QA (PWA)), and 
    the numerical one based on the SRB model in Eqs. \eqref{eq:SRB model}-\eqref{eq:SRB sub model 1} 
    (CFD (SRB)) is reported in Figs. \ref{fig:num ver fluid A}(c) and \ref{fig:num ver fluid B}(c).
    The quasi-analytical solution reproduces well the numerical one, reporting a
    slightly lower value about 2.5 \% at the inlet in the first case in Fig. \ref{fig:num ver fluid A}(c). 
    This outcome is due to the PWA of the rheological model which gives lower
    values of the axial pressure gradient in the final part of the pipe, which is characterized
    by a smaller hydraulic conductivity than the inlet and central regions with 
    higher radius values (see Eqs. \eqref{eq:hyd cond con flows F(r)}, \eqref{eq:hyd cond con flows}).
    Indeed, Fig. \ref{fig:num ver fluid A}(a)
    reports the working shear rate window at the outlet section calculated between
    $r=0.05R_{out}$ and $r=R_{out}$, clearly showing how the PWA
    returns a lower viscosity, and thus lower pressure values.
    The comparison of the velocity fields in Figs. \ref{fig:num ver fluid A}(d)-\ref{fig:num ver fluid A}(f)
    and Figs. \ref{fig:num ver fluid B}(d)-\ref{fig:num ver fluid B}(f) also shows
    a very good agreement between the quasi-analytical and numerical solutions, 
    for both types of fluid considered. 
    It is noteworthy the effects of higher shear-thinning properties in the 
    second case (fluid B) leading to a pressure drop of about 6.5 times lower 
    in Fig. \ref{fig:num ver fluid A}(c) and Fig. \ref{fig:num ver fluid B}(c), 
    a flatter axial velocity profile in Fig. \ref{fig:num ver fluid A}(e) and Fig. \ref{fig:num ver fluid B}(e) 
    and a radial velocity profile more shifted towards the pipe wall 
    in Fig. \ref{fig:num ver fluid A}(f) and Fig. \ref{fig:num ver fluid B}(f) 
    at the outlet section, than the first case (fluid A).
    
\clearpage
    
    \begin{figure}[!tb]
    	\centering
    	\subfloat[axis - pressure]{\includegraphics[width=.75\textwidth]{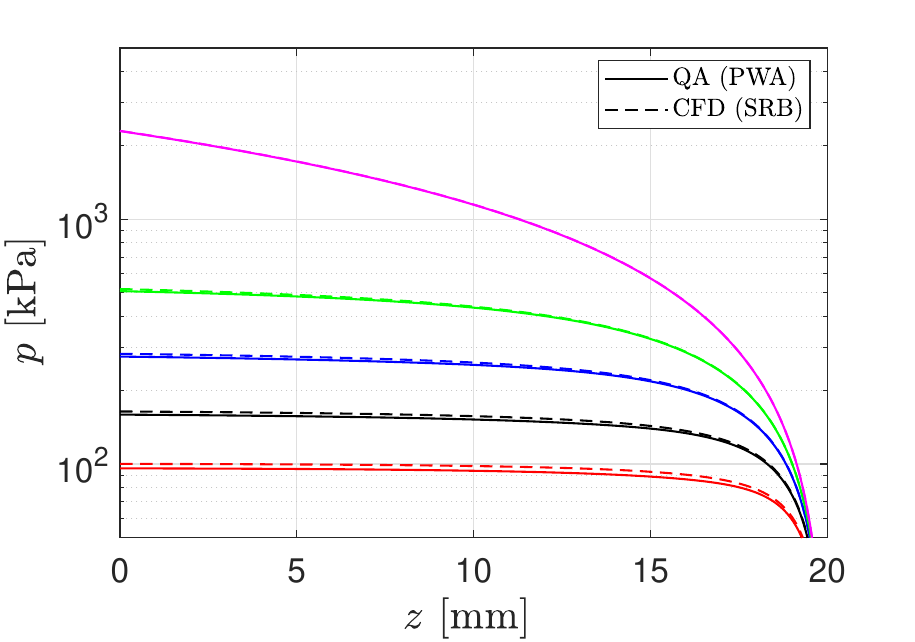}} \\
    	\subfloat[axis - axial velocity]{\includegraphics[width=.75\textwidth]{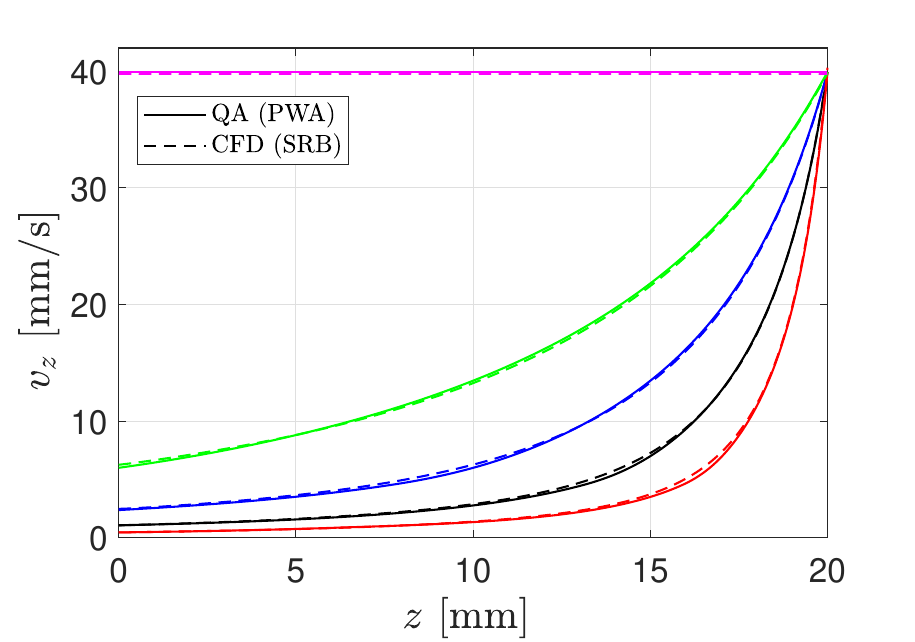}} \\
    	\caption{Numerical verification of the quasi-analytical solution (QA (PWA))
    	with comparison to the numerical solution (CFD (SRB)) by varying the taper angle.
    	$\theta = 0^{\circ}$ (magenta); $\theta = 1^{\circ}$ (green); $\theta = 2^{\circ}$ (blue);
    	$\theta = 3.58^{\circ}$ (black); $\theta = 6^{\circ}$ (red).}
    	\label{fig:num ver con angles}
    \end{figure}
    Next, another verification of the quasi-analytical solution has been carried 
    out in Fig. \ref{fig:num ver con angles} by varying the pipe taper angle of the 
    pipe from $\theta = 0^{\circ}$ (i.e. the cylindrical case, see Appendix A) 
    to $\theta = 6^{\circ}$, considering the fluid A in Table \ref{tab:fluid rheo param values}. 
    The inlet radius has been varied correspondingly while keeping the length 
    and the outlet radius constant.
    The pressure field along the pipe axis in Fig. \ref{fig:num ver con angles}(a) 
    reports a good agreement between the quasi-analytical solution (QA (PWA)) and 
    the numerical one (CFD (SRB)). As discussed before, the quasi-analytical
    solution shows a slightly lower value than the numerical one, ranging from 
    the 0.17 \% for the cylindrical case $\theta = 0^{\circ}$, up to about the 
    4.1 \% for the larger taper angle $\theta = 6^{\circ}$.
    The comparison of the axial velocity field along the axis in Fig. \ref{fig:num ver con angles}(b)
    shows almost overlapping solutions.
    
    \begin{figure}[!tb]
    	\centering
    	\subfloat[varying $\mu_0$]{\includegraphics[width=.45\textwidth]{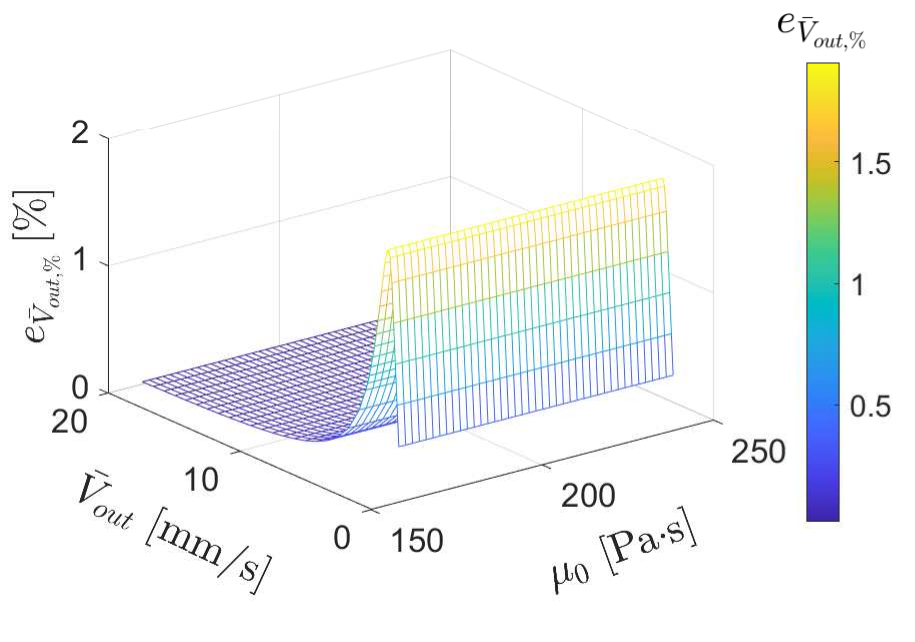}} \qquad
    	\subfloat[varying $\dot{\gamma}_0$]{\includegraphics[width=.45\textwidth]{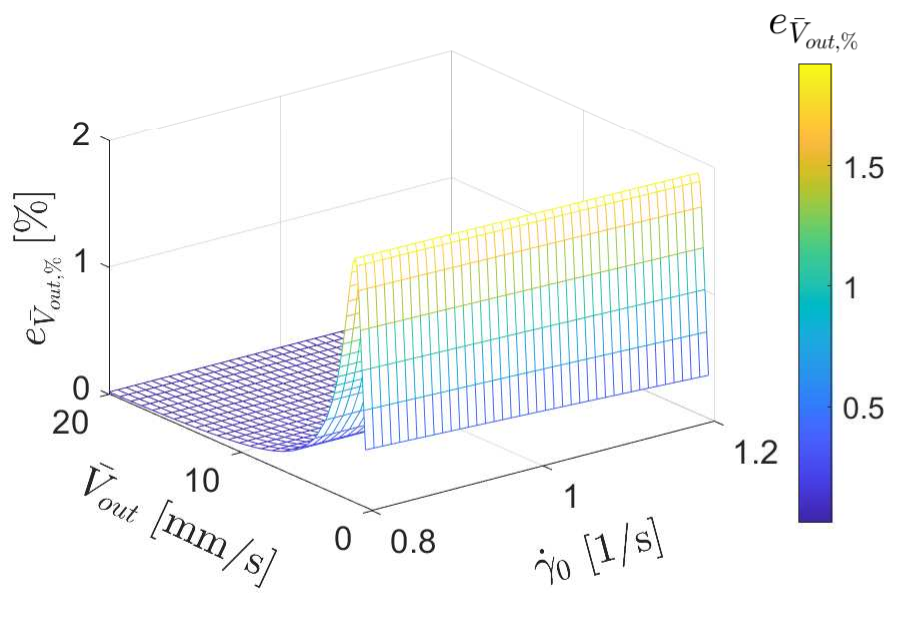}} \\
    	\subfloat[varying $n$]{\includegraphics[width=.45\textwidth]{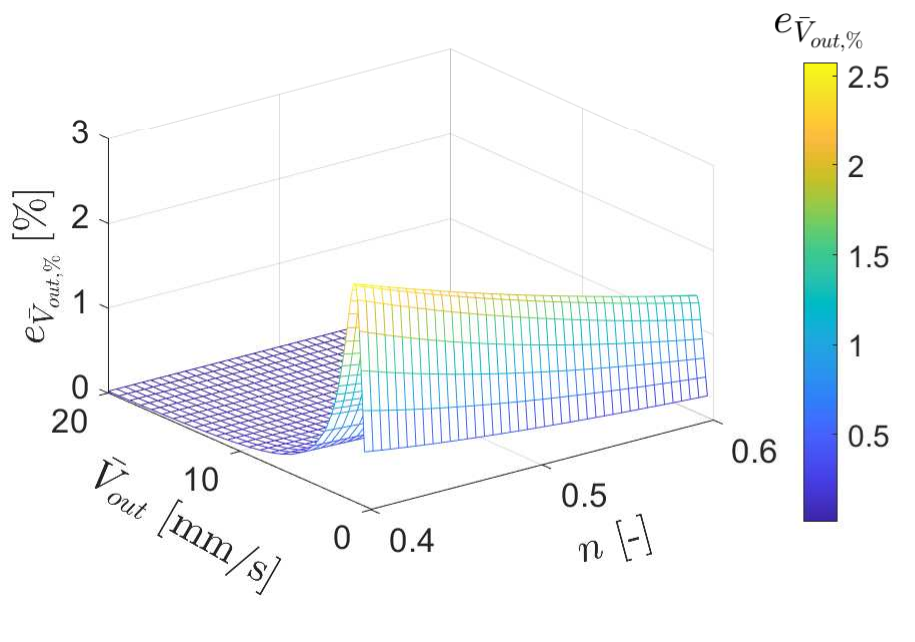}} \qquad
    	\subfloat[varying $\dot{\gamma}_\infty$]{\includegraphics[width=.45\textwidth]{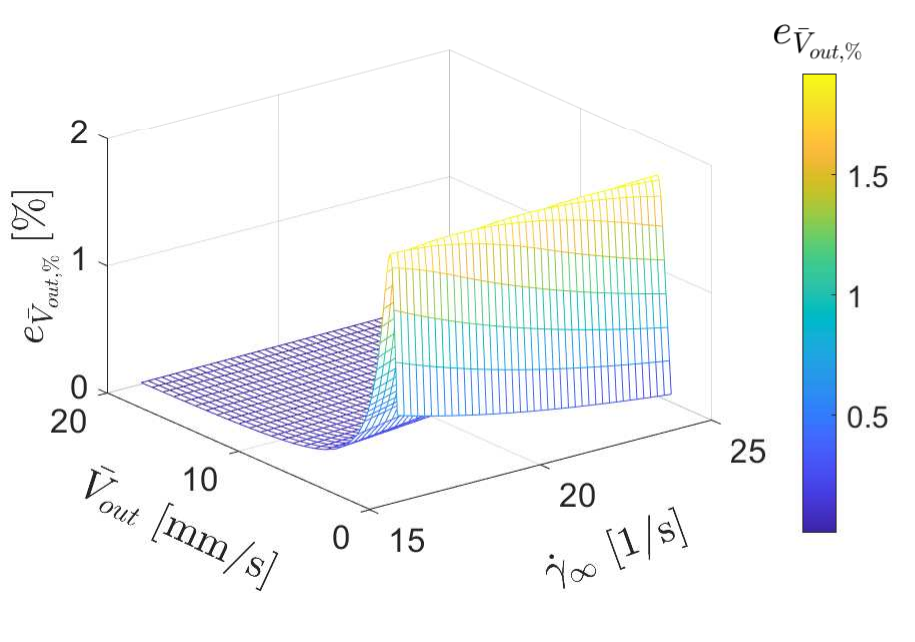}}
    	\caption{Parametric analysis of the mean extrusion velocity percentage error (Eq. (\ref{eq:errors num-ana sols}))
    	by varying  one by one the parameter values of fluid A in Table \ref{tab:fluid rheo param values} 
    	considered as nominal values. 
    	(a) varying $\mu_0$, (b) varying $\dot{\gamma}_0=1/\lambda_0$, (c) varying $n$, and (d) varying $\dot{\gamma}_\infty=1/\lambda_\infty$.}
    	\label{fig:para study num-ana error}
    \end{figure}
    Then, a further validation of the quasi-analytical solution has been performed
    through the parametric campaign shown in Fig. \ref{fig:para study num-ana error}.
    The analytical velocity profile at the outlet section has been compared with
    the numerical solution obtained by solving Eq. \eqref{eq:mom bal eqs con inelastic}
    for the SRB model (i.e. the non-approximated version of PWA) with the same
    $|p'|_{out}$, by analysing the mean extrusion velocity percentage error $e_{\bar{V}_{out,\%}}$
    between the analytical (PWA) and numerical (SRB) solutions
    \begin{equation}\label{eq:errors num-ana sols}
        e_{\bar{V}_{out,\%}} = \frac{\frac{1}{R_{out}^2}\int_{0}^{R_{out}}2r\left|v_{z,\text{PWA}}(r,z_{out})-v_{z,\text{SRB}}(r,z_{out})\ \right|dr}{\bar{V}_{out,\text{SRB}}}\ .
    \end{equation} 
    Each rheological parameter ($\mu_0, \lambda_0, \lambda_\infty, n$)
    of the fluid A reported in Table \ref{tab:fluid rheo param values} has been
    varied in a range of $\pm \, 20\, \%$ from its nominal value
    while keeping the others fixed with mean extrusion velocities ranging between
    $0-20$ mm/s.
    The maximum extrusion velocity error is about 1.9\%, except in the 
    parametric analysis of the power-law index $n$ which shows an error about
    2.6\%. For each analysis, the maximum error is reached in the region of the mean
    extrusion velocity close to $1.2$ mm/s, corresponding to the final zone
    of the MSR regime; then, the errors decrease as the mean extrusion velocity
    increases, moving the flow condition to the HSR regime.  
    Furthermore, it is noteworthy in Fig. \ref{fig:para study num-ana error}(c) 
    how the error decreases as the the power-law index increases, since the
    fluid tends to the Newtonian behaviour.

  \subsection{Flow analysis}
      \begin{table}[!htb]
          \caption{Values of rheological parameters for the SRB model ($a=2$) and PWA
          of the fluid analysed in Fig. \ref{fig:results:flow chart}.}
          \centering 
          \vspace{5pt}
         \begin{tabular}{ccc ccc  }
	  		\hline
            $\mu_{0}$               & $\mu_\infty$            & $\lambda_{0}$ & $\lambda_{\infty}$ & $n$   & $K$          \\ [0.5ex]
            $[\text{Pa $\cdot$ s}]$ & $[\text{Pa $\cdot$ s}]$ & $[\text{s}]$  & $[\text{s}]$       & $[-]$ & [Pa$\cdot$s$^{n}$]  \\ [0.5ex]
            \hline \vspace{-12pt}  \\
            $200$                   &  $8.94$                 & $5$           & 0.01               & 0.5   & $89.4$ \\[0ex]
	  		\hline
         \end{tabular}
         \label{tab:res:rheo values flow analysys}
    \end{table}
    \begin{figure}[!tb]
                \centering
                \includegraphics[width=0.9\textwidth]{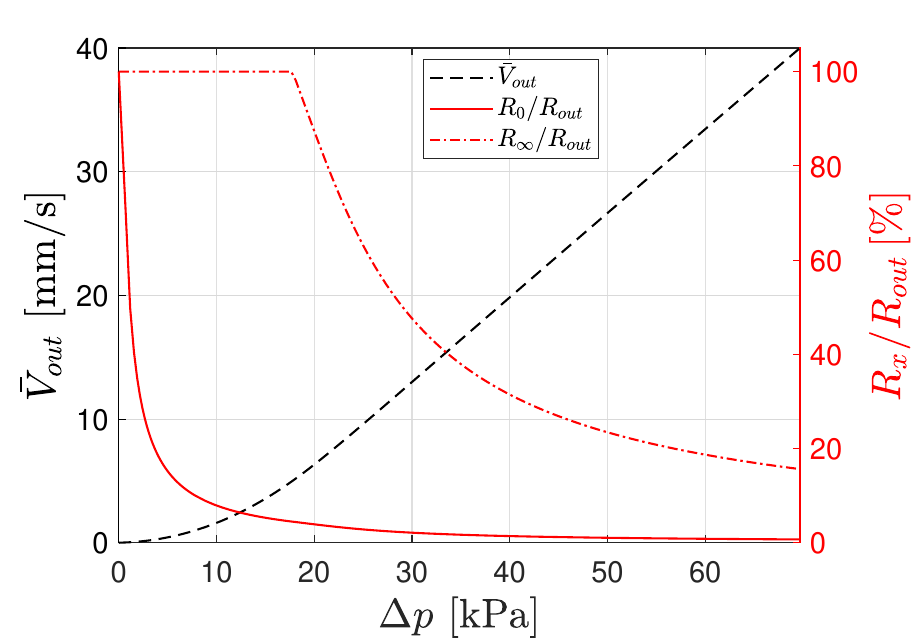}\hspace{+0.0cm}
                \caption{Mean extrusion velocity at the pipe outlet as a function of the
                applied pressure drop (black dashed line, left axis); corresponding $R_0$ and
                $R_\infty$ values in Eqs. \eqref{eq:R0 and R infinity con} (red continuous and dash-dotted lines, right axis).}
                \label{fig:results:flow chart}
    \end{figure}
    The solution in Eqs. \eqref{eq:PWM-LSR expressions conical}-\eqref{eq:PWM-HSR expressions conical}
    allows a direct assessment of the pressure drop applied to the nozzle and of
    the corresponding mean extrusion velocity, together to the evaluation of the
    Newtonian/power-law conditions of the flow.
    It has been applied to an inelastic fluid described through the SRB model
    in Eq. \eqref{eq:SRB model} with the rheological values reported in Table \ref{tab:res:rheo values flow analysys}.
    The mean extrusion velocity at the outlet section $ \bar{V}_{out}=Q/(\pi R_{out}^2)$
    calculated from the flow rate in Eqs. \eqref{eq:PWM-LSR expressions conical}-\eqref{eq:PWM-HSR expressions conical}
    is shown in Fig. \ref{fig:results:flow chart} (left axis) as a function of the
    pressure drop $\Delta p$ applied to the nozzle. 
    Moreover, Fig. \ref{fig:results:flow chart} (right axis) shows the $R_0$ and 
    $R_\infty$ functions in Eqs. \eqref{eq:R0 and R infinity con} evaluated at
    the outlet and corresponding to the applied $\Delta p$.
    It is possible to note how the flow immediately exits from the LSR condition 
    at very low pressure drop values. Indeed, $R_0$ in Eq. \eqref{eq:R0 and R infinity con} 
    quickly drops down.
    This behaviour is due to the low value for $1/\lambda_0$ and the
    geometry at hand, which determine $|{\dot{\gamma}}_{wall}|>1/\lambda_0$ at
    very low values of $\Delta p$, causing the flow to leave the LSR regime.
    Then, the flow is characterized by a MSR regime up to a mean extrusion velocity of
    about $5$~mm/s, and enters in the HSR regime.
    By investigating the slope of the mean extrusion velocity curve, it is interesting to
    see the linear relationship between $\bar{V}_{out}$ and $\Delta p$ in the HSR flow 
    condition referring to the Newtonian relation in Eq. \eqref{eq:con newto:flow rate}, 
    and how the shear-thinning properties of the fluid in the MSR flow condition 
    (i.e. until $R_0/R_{out} < 100\%$ and $R_\infty/R_{out} = 100\%$) allow to 
    increase the mean extrusion velocity slope compared to the initial Newtonian $\mu_0$
    zone.  
  
  \subsection{Damage wall shear stress}\label{subsec:res:damage wss}
    \begin{figure}[!tb]
    	\centering
    		\subfloat[varying $\mu_0$ and $\dot{\gamma}_0$; $\tau_{dam.}=100$ Pa. ]{\includegraphics[width=.45\textwidth]{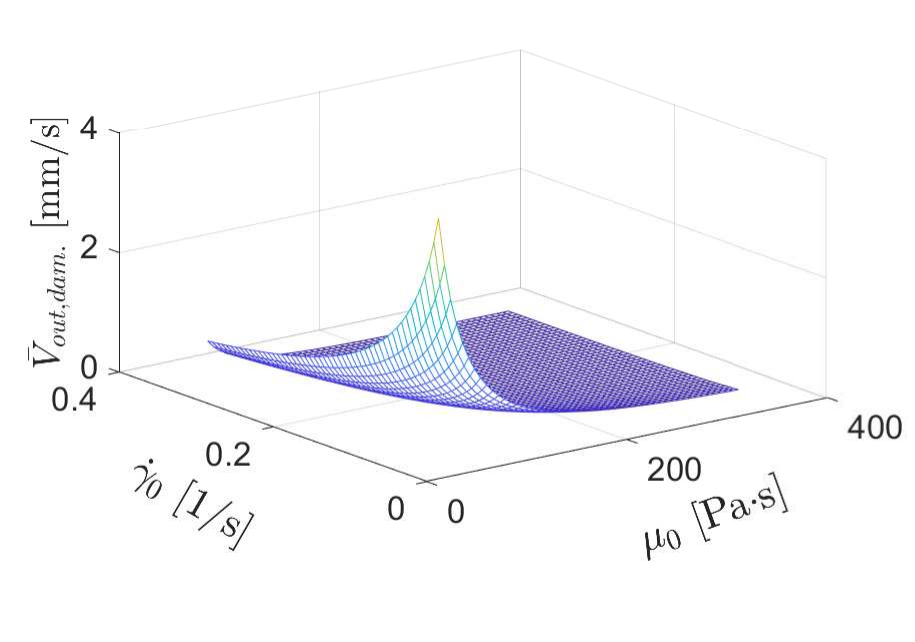}}\qquad
    		\subfloat[varying $n$ and $\dot{\gamma}_\infty$; $\tau_{dam.}=100$ Pa.]{\includegraphics[width=.45\textwidth]{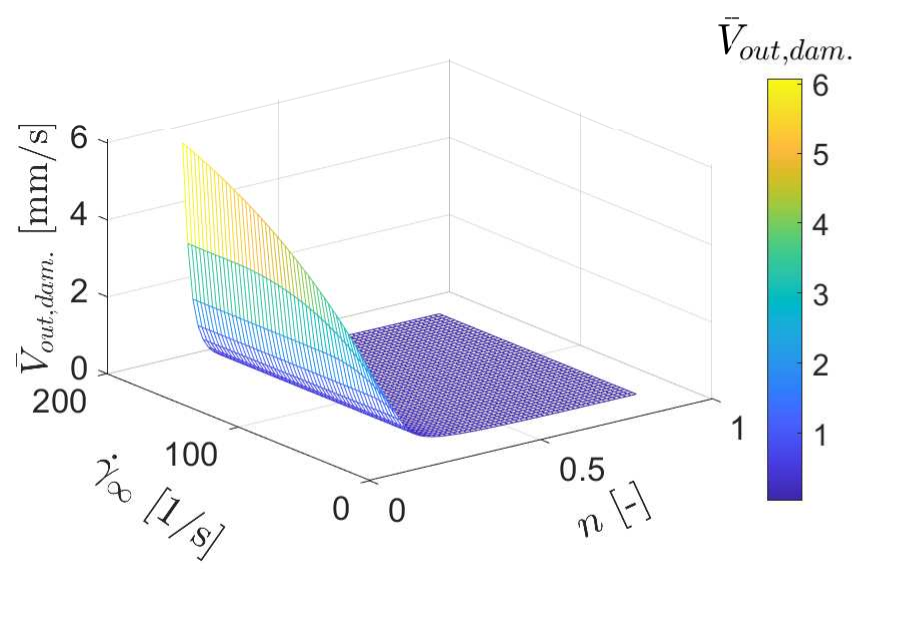}} \\
    		\subfloat[varying $\mu_0$ and $\dot{\gamma}_0$; $\tau_{dam.}=500$ Pa. ]{\includegraphics[width=.45\textwidth]{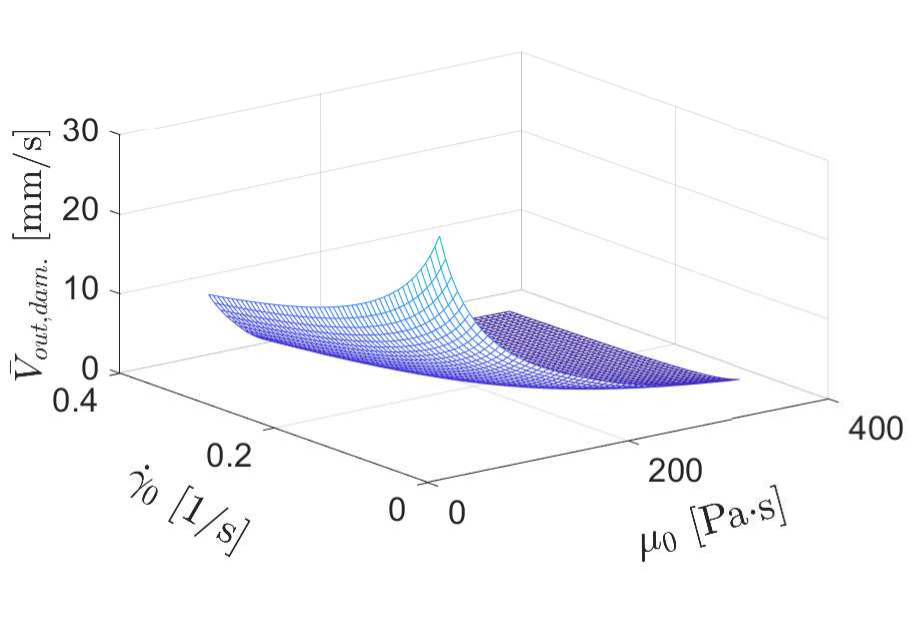}}\qquad
    		\subfloat[varying $n$ and $\dot{\gamma}_\infty$; $\tau_{dam.}=500$ Pa.]{\includegraphics[width=.45\textwidth]{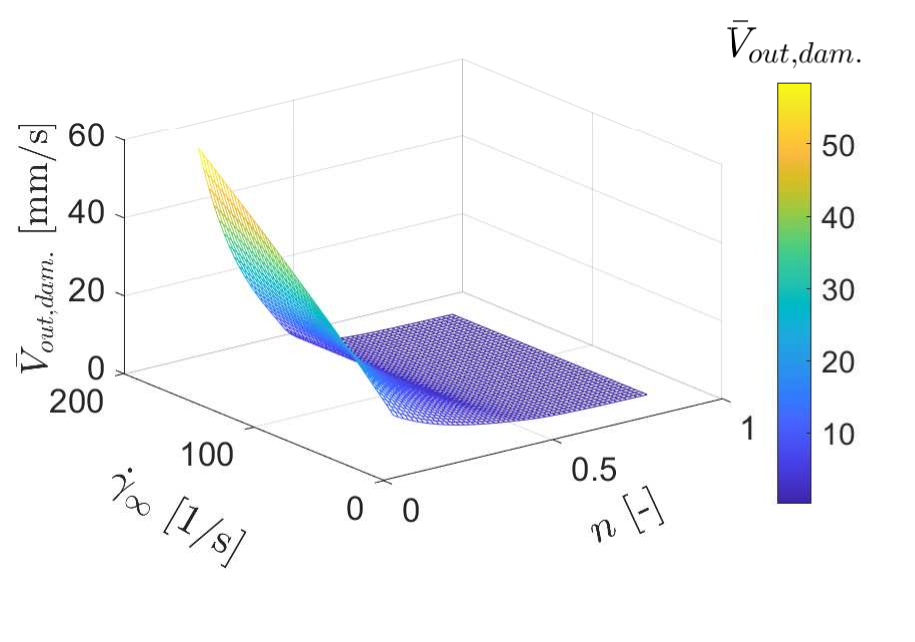}}
    	\caption{Parametric analyses of the damage mean extrusion velocity at the outlet
    	(Eq. (\ref{eq:crit. extr. vel.})) by varying the parameter values in Table \ref{tab:res:rheo values flow analysys}
    	considered as nominal values, for two cell damage shear stress.
    	(a,b) $\tau_{dam.}=100$ Pa, (c,d) $\tau_{dam.}=500$ Pa;
    	(a),(c) varying $\mu_0$ and $\dot{\gamma}_0=1/\lambda_0$,
    	(b),(d) varying $n$ and $\dot{\gamma}_\infty=1/\lambda_\infty$.}
    	\label{fig:para analysis crit. extr. vel.}
    \end{figure}
    In many applications it is often desirable to evaluate the shear stress
    within the flow and its maximum value occurring at the nozzle wall namely the
    wall shear stress (see Eq. \eqref{eq:shear stress con inelastic}),
    such as in the study of cardiovascular diseases \cite{katritsis2007wall, cecchi2011role},
    or in extrusion bioprinting \cite{boularaoui2020overview, conti2022models}.
    In the latter case a key requirement is to limit the shear stresses,
    since they may cause cell damages \cite{nair2009characterization, shi2018shear, chirianni2024development}.
    The SRB model and PWA allow to design a safety operating condition to limit
    the wall shear stress during the extrusion.
    For example, by taking as reference the infinity-shear stress $\tau_\infty=\mu_\infty \dot{\gamma}_\infty$
    of the SRB model, considering a damage shear stress of the cell $\tau_{dam.}=\phi \tau_\infty$
    with $\phi \gtreqless 1$ corresponding to the the scalar ratio
    between $\tau_{dam.}$ and $\tau_{\infty}$, through the SRB model and PWA
    a safety operating condition could be to ensure a wall shear stress at the
    outlet section lower than the damage shear stress of the cell, reading
    \begin{equation}\label{eq:crit. extr. vel. B}
          \tau_{max}=\tau_{out, wall}=\frac{R_{out}}{2} \left| \frac{d p}{d z} \right|_{out}  \le \tau_{dam.}=\phi \tau_\infty=\phi \mu_0{\dot{\gamma}}_0\left(\frac{{\dot{\gamma}}_\infty}{{\dot{\gamma}}_0}\right)^n\ .
    \end{equation}
    Eq. \eqref{eq:crit. extr. vel. B} establishes a relation between
    the operating conditions $p'$ (and implicitly $\Delta p$
    and $Q$), the needle geometry $R_{out}$ (and implicitly $L$ and $\theta$),
    the rheological properties of the fluid ($\mu_0, \lambda_0, \lambda_\infty, n$),
    and the damage shear stress of the cell ($\phi= \tau_{dam.}/\tau_{\infty}$).
    The corresponding damage mean extrusion velocity at the outlet $\bar{V}_{out, dam.}$
    can be computed from the flow rate in Eqs. \eqref{eq:PWM-LSR expressions conical}-\eqref{eq:PWM-HSR expressions conical}
    depending on the value of $\phi$, reading
    \begin{equation}\label{eq:crit. extr. vel.}
        \bar{V}_{out, dam.} = \frac{ Q(|p'|_{out, dam.}) }{\pi R_{out}^2}\ , \qquad \left| p' \right|_{out,dam.} = \frac{2\phi \mu_0{\dot{\gamma}}_0}{R_{out}} \left(\frac{{\dot{\gamma}}_\infty}{{\dot{\gamma}}_0}\right)^n\ .
    \end{equation}

    As application example, two parametric analyses has been carried out in Fig. \ref{fig:para analysis crit. extr. vel.}
    for the damage mean extrusion velocity at the outlet $\bar{V}_{out,dam.}$ in Eq. \eqref{eq:crit. extr. vel.},
    by imposing a $\tau_{dam.} = \tau_{out, wall} = 100$ Pa in Figs. \ref{fig:para analysis crit. extr. vel.} (a-b),
    and a $\tau_{dam.} = \tau_{out, wall} = 500$~Pa
    in Figs. \ref{fig:para analysis crit. extr. vel.} (c-d).
    Then, the parameters have been varied in pairs in a range of $\pm \, 75\%$ 
    from their nominal values in Table \ref{tab:res:rheo values flow analysys},  
    while keeping the others fixed.
    It is noteworthy how decreasing the zero-shear rate viscosity $\mu_0$, the 
    zero-shear rate $\dot{\gamma}_0 = 1/\lambda_0$ and the power-law index $n$, and increasing
    the infinity-shear rate $\dot{\gamma}_\infty=1/\lambda_\infty$ allows greater values of
    extrusion velocities without the occurence of damage for both cases.
    This outcome arises since the flow would tend to change from an high viscosity 
    LSR regime, to a shear-thinning MSR regime.
    Furthermore, it is possible to note in the second case with higher damage
    shear stress value, how the damage extrusion velocity can reach a value around
    ten times higher than in the first case.

\section{Conclusions}
  In this work, a novel quasi-analytical solution for ``Carreau-Yasuda-like'' fluids
  flowing in slightly tapered pipes, and whose viscosity rheological response is
  modelled with a PWA has been presented.
  In particular, the PWA is characterized by two constant plateau at low and high
  shear rates connected by a shear-thinning branch.
  In literature, several solutions have been reported since the last century 
  considering both a Newtonian \cite{blasius1910laminare, forrester1970flow, langlois1972creeping, kotorynski1995viscous, sisavath2001creeping} and a non-Newtonian 
  inelastic \cite{priyadharshini2015biorheological, fusi2020flow} fluids. 
  In particular, Priyadharshini and Ponalagusam \cite{priyadharshini2015biorheological} 
  and Fusi \etal \cite{fusi2020flow} provided solutions for yield stress fluids, 
  described through the Hershel-Bulkley and Bingham model, respectively. 
  However, these models do not assume viscosity plateaus at high and low shear 
  rates.
  Quasi-analytical solutions of the main flow problem variables for conical nozzles
  have been provided in Eqs. \eqref{eq:PWM-LSR expressions conical}-\eqref{eq:PWM-HSR expressions conical}.
  Morover, Algorithm \ref{alg:dpdz PWM} enables the evaluation of the axial
  pressure gradient required for the solution computation.
  The solutions have been verified through numerical procedures
  showing a robust consistency of the analytical approach, with a maximum error
  on the computed velocity and pressure profiles well below engineering
  applications (see Figs. \ref{fig:num ver fluid A}-\ref{fig:para study num-ana error}).
  Next, the proposed quasi-analytical framework allows fast and
  ready-to-use screening evaluations on the mutual impact of main process
  variables on the flow dynamics of non-Newtonian inelastic fluids
  (see Fig. \ref{fig:results:flow chart}).
  For instance, an exemplary application of the quasi-analytical solution
  for the shear stresses evaluation in extrusion bioprinting has
  been reported in Fig. \ref{fig:para analysis crit. extr. vel.}.
  In addition, this approach could support the verification of new numerical
  approaches for GNFs.
  It is noteworthy that this framework is not limited to cone-shaped pipes, 
  but in general is valid for slightly tapered conducts with a small decrease 
  of radius $dR/dz$ along the axis, as predicted by lubrication theory \cite{bird1987dynamics}.
  However, generally pipes can present more complex geometries and, even 
  remaining in the case of axisymmetric channels, a numerical computational fluid
  dynamics approach could be required.
  Indeed, nozzles with a non-small taper angle, or with sharp and local flow 
  section reductions, present non-negligible radial components in the mass and 
  momentum conservation equations, and cannot be treated in terms of pressure 
  and velocity solutions through an analytical method \cite{ferziger2002computational}.

\section*{Acknowledgments}

Part of this work was carried out with the support from the Italian National Group for Mathematical Physics GNFM-INdAM.

Michele Marino and Giuseppe Vairo acknowledge financial support by the Italian Ministry of University and Research (MUR) under the National Recovery and Resilience Plan (NRRP), PRIN 2022 program, Project 2022T3SLAZ – CUP E53D23003700006.

\newpage
\section*{}
\bibliography{mybibfile}


\appendix

\section{Conical flow of a Newtonian fluid}
    In the case of a Newtonian fluid with a constant viscosity $\mu$,
    an order-of-magnitude analysis \cite{bird1987dynamics} from the mass
    conservation Eq. \eqref{eq:mass cons eq} yields the following relationship
    between the reference axial $V_z$ and radial $V_r$ velocities
    \begin{equation}\label{eq:magn analysis 2 conical flow}
      V_r = V_z \frac{R_{out}}{L}\left( 1 - \chi^2 \right)  \ ,
    \end{equation}
    where $V_z$ and $V_r$ are the reference axial and radial velocity values
    respectively, and $\chi=R_{out}/R_{in}$.
    Next, from the momentum balance Eqs. \eqref{eq:mom cons eq generic conical}
    by neglecting the smaller terms the comparison of the inertial to the viscous
    terms gives the following relationship
    \begin{equation}\label{eq:conical cond newto 1}
          \frac{ O\left( \text{inertial terms} \right) }{O\left( \text{viscous terms} \right) } = Re \frac{R_{out}}{L} \left( 1-\chi^2\right) \ ,
    \end{equation}
    with the Reynolds number defined as $Re=\rho V_z R_{out}/\mu$.

    In case of low Reynolds number flows, the previous equation indicates the
    inertial terms are negligible than the viscous ones. But it is worth nothing
    that even with non-negligible Reynold numbers, the geometrical factor
    $R_{out}(1-\chi^2)/L$ ensures the negligibility of the inertial terms.
    Furthermore, in case of a cylindrical pipe ($\chi=1$) it returns that
    the inertial terms are identically null.

    Next, the comparison of the pressure gradients terms turns out
    \begin{equation}\label{eq:conical cond newto 2}
          \frac{ O\left( \derpar{p}{r} \right) }{ O\left( \derpar{p}{z} \right) } = \frac{ R_{out} }{ {L} } \left( 1- \chi^2 \right) \ ,
    \end{equation}
    showing that in case of a slightly tapered pipe also the radial pressure
    gradient can be neglected.
    Thus, by neglecting the smaller terms the momentum balance Eq. (\ref{eq:mom cons eq generic conical}b)
    results
    \begin{equation}\label{eq:mom bal eqs con}
           \derpar{p}{z} \simeq \frac{d p}{d z}
           \simeq \frac{1}{r}\derpar{}{r} \left(  r \mu \derpar{v_z}{r} \right)
           = \mu \left[ \frac{1}{r}\derpar{}{r} \left(  r\derpar{v_z}{r} \right) \right]  \ ,
    \end{equation}
    which is the equivalent form of the momentum balance equation for a cylindrical tube \cite{bird1987dynamics}.
    The integration of the Eq. \eqref{eq:mom bal eqs con} in an arbitrary
    axial section of the pipe with the application of
    the symmetric flow boundary condition (i.e. $\partial v_z/\partial r |_{r=0} =0$)
    and the no-slip boundary condition at the wall (i.e. $v_z(r=R(z),z)=0$)
    gives the flow rate equation
    \begin{equation}\label{eq:flow rate eq con newto}
          Q = -\frac{d p}{dz}(z)\frac{\pi R^4(z)}{8\mu},
    \end{equation}
    which is equivalent to the Hagen-Poiseuille law for cylindrical flows,
    but with the substantial difference of a non-constant radius.
    Therefore, also the pressure gradient varies along the axis in order to
    satisfy the flow conservation in each section.
    Thus, given a flow rate value, from the previous equation and by knowing
    the geometric function describing the cross section variation along the
    axis in Eq. \eqref{eq:radius conical nozzle}, the pressure solution results
    \begin{equation}\label{eq:press conical newto}
        \begin{aligned}
          p(z) &= p_{in} -\frac{Q 8\mu }{\pi 3\theta} \left[\frac{1}{(R_{in} - \theta z)^3} - \frac{1}{R_{in}^3}\right] \\
               &= p_{in} -\frac{Q 8\mu }{\pi R_{in}^4} z \left[1 + \frac{2\theta z}{R_{in}} + \sum_{m=3}^\infty \binom{-3}{m} \frac{(-1)^m}{3} \left( \frac{\theta z}{R_{in}}\right)^{m-1} \right],
        \end{aligned}
    \end{equation}
    where $p_{in}$ is the pressure value at the pipe inlet.
    In the third member assuming a $|\theta z/R_{in}|<1$ the Taylor
    expansion has been applied to highlight the conical contribution
    with respect to the cylindrical case (with a linear variation of
    the pressure), which vanishes for a null taper angle $\theta = 0^\circ$.

    Next, the velocity, shear rate, shear stress and flow rate solutions
    turn out
    \begin{align}
          & v_z(r,z) = \frac{2Q}{\pi R^2(z)} \left[ 1 - \left( \frac{r}{R(z)} \right)^2 \right]
                  = 2\bar{V}(z)\left[ 1 - \left( \frac{r}{R(z)} \right)^2 \right], \label{eq:ax vel newto} \\[10pt]
          & v_r(r,z) = -\frac{2Q}{\pi R^2(z)} \left[ 1 - \left( \frac{r}{R(z)} \right)^2 \right] \frac{\theta r}{R(z)}
                  = -v_z(r,z) \frac{\theta r}{R(z)} , \label{eq:rad vel newto} \\[10pt]
          & \dot{\gamma}_{zr}(r,z) \simeq \dot{\gamma}(r,z) = - \frac{4Qr}{\pi R^4(z)} = - \frac{4\bar{V}(z)r}{R^2(z)},  \label{eq:cap3QAM:conical shear rate newto}  \\[10pt]
          & \tau_{zr}(r,z) \simeq \tau(r,z) = - \frac{4Q\mu r}{\pi R^4(z)} = - \frac{4\bar{V}(z)\mu r}{R^2(z)},  \\[10pt]
          & Q = -\frac{d p}{dz}(z)\frac{\pi R^4(z)}{8\mu} = \frac{\Delta p}{L}\frac{3\pi\left(R_{in}-R_{out}\right)}{8\mu\left(\frac{1}{R_{out}^3}-\frac{1}{R_{in}^3}\right)}
          = \frac{\Delta p}{L}\frac{\pi R_{in}^4}{8\mu}\left( 1 - \frac{1+\chi+\chi^2-3\chi^3 }{1+\chi+\chi^2} \right) \label{eq:con newto:flow rate},
    \end{align}
    where $\bar{V}(z)$ is the mean velocity in the specific axial section.
    The radial velocity solution in Eq. \eqref{eq:rad vel newto} derives from
    the integration of the mass conservation in Eq. \eqref{eq:rad vel - mass cons eq},
    with the application of the boundary condition at the pipe axis $v_r(r=0, z)=0$.
    It is noteworthy that the radial velocity solution nulls out in case of
    a cylindrical tube (i.e. $\theta=0^\circ$).


\section{Conical flow of a power-law fluid}
    By performing the same order-of-magnitude analysis and given the same assumptions
    of the Newtonian flow in Appendix A, by considering a power-law fluid
    the magnitude relation in Eq. \eqref{eq:magn analysis 2 conical flow} is
    still valid.
    Therefore, the corresponding strain rate tensor results
    \begin{equation}
            \vect{E} =
            \begin{bmatrix}
                E_{rr}      &  0                  &  E_{zr} \\
                            &  E_{\theta \theta}  &  0 \\
                \text{sym}  &                     &  E_{zz}
            \end{bmatrix}
            =
            \begin{bmatrix}
                \derpar{v_r}{r}  &  0           &   \dfrac{1}{2} \left( \derpar{v_r}{z} + \derpar{v_z}{r} \right) \\
                            &  \dfrac{v_r}{r}  &   0 \\
                \text{sym}  &                   & \derpar{v_z}{z}
            \end{bmatrix},
    \end{equation}
    with a order-of-magnitude analysis equal to
    \begin{equation}
            O\left( \vect{E}\right) =
            \begin{bmatrix}
                O\left(\dfrac{V_z(1-\chi^2)}{L} \right)  &  0                                            &  O\left(\dfrac{V_z}{R_{out}} \right) \\
                                                             &  O\left(\dfrac{V_z(1-\chi^2)}{L} \right)  &  0 \\
                \text{sym}                                   &                                               &  O\left(\dfrac{V_z(1-\chi^2)}{L} \right)
            \end{bmatrix} .
    \end{equation}
    By comparing each components, it is possible to note how the
    greatest term is $E_{zr} \simeq \partial v_z/\partial r$,
    which allows to approximate the strain rate tensor norm in Eq. \eqref{eq:strain rate magn}
    as
    \begin{equation}
          \dot{\gamma} \simeq \left| \derpar{v_z}{r} \right| .
    \end{equation}
    non-Newtonian fluid described by the power-law model \cite{waele1923viscometry, ostwald1925ueber}
    \begin{equation}
            \mu(\dot{\gamma}) = K \dot{\gamma}^{n-1}\ ,
    \end{equation}
    leads to the corresponding stress tensor
    \begin{equation}
            \vectb{\tau} = 2 \mu( \dot{\gamma}) \vect{E} =2 K \dot{\gamma}^{n-1} \vect{E} =
        \begin{bmatrix}
                \tau_{rr}   &  0                     &  \tau_{zr} \\
                            &  \tau_{\theta \theta}  &  0 \\
                \text{sym}  &                        &  \tau_{zz}
        \end{bmatrix}\ .
    \end{equation}
    The corresponding order-of-magnitude analysis equals to
    \begin{equation}
        O\left( \vectb{\tau}\right) =
        \begin{bmatrix}
                O\left(\zeta\right)  &  0                    &  O\left(K \left( \dfrac{V_z}{R_{out}} \right)^n \right) \\
                                     &  O\left(\zeta\right)  &  0 \\
                \text{sym}                                   &    &  O\left(\zeta\right)
        \end{bmatrix} ,
    \end{equation}
    where $\zeta = K \left( \dfrac{V_z}{R_{out}} \right)^n \dfrac{R_{out}(1-\chi^2)}{L}$.

    It is worth nothing that how for a null taper angle
    (i.e. $\theta = 0^\circ,\ \chi = 1$), the diagonal components of
    the strain rate and stress tensors vanish as in the cylindrical flow case.
    By performing a order-of-magnitude analysis of the momentum balance Eqs. \eqref{eq:mom cons eq generic conical},
    the comparison of the inertial to the viscous terms gives the following
    relationship
    \begin{equation}\label{eq:conical cond powlaw 1}
          \frac{ O\left( \text{inertial terms} \right) }{O\left( \text{viscous terms} \right) } = Re_{(K,n)} \frac{R_{out}}{L} \left( 1-\chi^2\right) \ ,
    \end{equation}
    where $Re_{(K,n)} = \rho V_z^{2-n} R_{out}^n/K$ is the Reynolds number specified for a power-law fluid
    as reported in \cite{dodge1959turbulent}.
    Next, the comparison of the pressure gradients terms gives the same
    previous relation in Eq. \eqref{eq:conical cond newto 2}
    \begin{equation}\label{eq:conical cond powlaw 2}
          \frac{ O\left( \derpar{p}{r} \right) }{ O\left( \derpar{p}{z} \right) } = \frac{ R_{out} }{ {L} } \left( 1- \chi^2 \right) \ .
    \end{equation}
    Thus, also for a power-law fluid when the Reynolds numbers
    and the geometrical factor $R_{out}(1-\chi^2)/L$ are small,
    it is possible to neglect
    the inertial terms  and the radial pressure gradient
    in the momentum balance equations,
    as for the previous Newtonian case.
    Therefore, the momentum balance Eq. (\ref{eq:mom cons eq generic conical}b)
    turns out
    \begin{equation}\label{eq:cap3QAM:mom bal eqs con powlaw}
           z)      \qquad  \derpar{p}{z} \simeq \frac{d p}{d z}  \simeq \frac{1}{r}\derpar{}{r}\left( r K\dot{\gamma}^{n-1} \derpar{v_z}{r}\right) \ .
    \end{equation}
    The integration of the Eq. \eqref{eq:cap3QAM:mom bal eqs con powlaw} in an arbitrary
    axial section of the pipe with the application of
    the symmetric flow boundary condition (i.e. $\partial v_z/\partial r |_{r=0} =0$)
    and the no-slip boundary condition at the wall (i.e. $v_z(r=R(z),z)=0$)
    gives the flow rate equation

    \begin{equation}\label{eq:flow rate eq con powlaw}
          Q = \left( -\frac{d p}{dz}(z) \frac{1}{2K} \right)^{\frac{1}{n}} \frac{\pi R^\beta (z)}{\beta}  ,
    \end{equation}
    where $\beta = (3n+1)/n$.
    Given a flow rate value and by knowing the geometric function describing
    the cross section variation of the pipe along the axis in
    Eq. \eqref{eq:radius conical nozzle}, from the previous equation
    the pressure solution results
    \begin{equation}
        \begin{aligned}
          p(z) &= p_{in} - \left( \frac{Q\beta }{\pi } \right)^n \frac{2K}{3n \theta }  \left[ \frac{1}{(R_{in} - \theta z)^{3n}} - \frac{1}{R_{in}^{3n}}\right] \\
               &= p_{in} - \left( \frac{Q\beta }{\pi } \right)^n \frac{2K}{R_{in}^{3n+1} } z \left[1 + \frac{(3n+1)\theta z}{2R_{in}} +
               \sum_{m=3}^\infty \binom{-3n}{m} \frac{(-1)^m}{3n} \left( \frac{\theta z}{R_{in}}\right)^{m-1} \right],
        \end{aligned}
    \end{equation}
    where in the third member assuming a $|\theta z/R_{in}|<1$ the Taylor
    expansion has been applied to highlight the conical contribution
    with respect to the cylindrical case (with a linear variation of
    the pressure), which vanishes for a null taper angle $\theta = 0^\circ$.
    Then, the velocity, shear rate, shear stress and flow rate solutions turn out
    \begin{align}
           & v_z(r,z) = \frac{\beta Q}{\alpha \pi R^2(z)} \left[ 1 - \left( \frac{r}{R(z)} \right)^\alpha \right]
                  = \frac{\beta}{\alpha} \bar{V}(z)\left[ 1 - \left( \frac{r}{R(z)} \right)^\alpha \right], \label{eq:ax vel powlaw}\\[10pt]
           & v_r(r,z) = -\frac{\beta Q}{\alpha \pi R^2(z)} \left[ 1 - \left( \frac{r}{R(z)} \right)^\alpha \right] \frac{\theta r}{R(z)}
                  = - v_z(r,z) \frac{\theta r}{R(z)} , \label{eq:rad vel powlaw} \\[10pt]
           & \dot{\gamma}_{zr}(r,z) \simeq \dot{\gamma}(r,z) = - \frac{\beta Q}{\pi }\left( \frac{r}{R^{3n+1}(z)} \right)^{\frac{1}{n}}
           = - \beta \bar{V}(z)\left( \frac{r}{R^{n+1}(z)} \right)^{\frac{1}{n}},   \label{eq:shear rate con powlaw} \\[10pt]
           & \tau_{zr}(r,z) \simeq \tau(r,z) = - \left(\frac{\beta Q}{\pi }\right)^n \frac{Kr}{R^{3n+1}(z)}
           = - \left(\beta \bar{V}(z)\right)^n \frac{Kr}{R^{n+1}(z)},  \\[10pt]
           Q &= \left( -\frac{d p}{dz}(z) \frac{1}{2K} \right)^{\frac{1}{n}} \frac{\pi}{\beta} R^{\beta}(z) = \left( \frac{\Delta p \, 3n}{2LK} \right)^{\frac{1}{n}} \frac{\pi}{\beta} \left( \frac{R_{in}-R_{out} }{\frac{1}{R_{out}^{3n}}-\frac{1}{R_{in}^{3n}}} \right)^{\frac{1}{n}}\\
           & =  \left( \frac{\Delta p }{2LK} \right)^{\frac{1}{n}} \frac{\pi R_{in}^\beta}{\beta} \left( 1 - \frac{1-\chi^{3n} -3n(1-\chi)\chi^{3n} }{1-\chi^{3n}} \right) \nonumber,
    \end{align}
    where $\alpha = (n+1)/n$, $\beta = (3n+1)/n$ and $\bar{V}(z)$ is the mean
    velocity in the specific axial section. It is interesting to note
    how the shear rate expression in Eq. \eqref{eq:shear rate con powlaw}
    evaluated at the pipe wall (i.e. for $r=R(z)$)
    recurs in the simplest case of the wall shear rate for a cylindrical duct
    as reported in \cite{rahimnejad2020rheological}.
    The radial velocity solution in Eq. \eqref{eq:rad vel powlaw} derives from
    the integration of the mass conservation in Eq. \eqref{eq:rad vel - mass cons eq},
    by applying the boundary condition at the pipe axis $v_r(r=0, z)=0$.
    It is noteworthy that the radial velocity solution nulls out in case of
    a cylindrical pipe (i.e. $\theta=0^{\circ}$).


\section{Integration functions of the quasi-analytical solution}
The integration functions of the quasi-analytical solution in Section \ref{sec:quasi-analytical sol}
read:
\begin{itemize}
  \item
  Medium-shear rate (MSR) flow, $\dot{\gamma}_0<\left|\dot{\gamma}_{wall}(z)\right|\le\dot{\gamma}_\infty, \tau_0<\left|\tau_{wall}(z)\right|\le\tau_\infty$,
  \begin{fleqn}[\parindent]

    \text{for} $r \le R_0(z)$,
    \begin{align}
        &\quad A_0(z) ={\frac{1}{\alpha}\left(-\frac{p'(z)}{2K}\right)}^\frac{1}{n}\left(R^\alpha(z)-R_0^\alpha(z)\right)\ -\ \frac{p'(z) R_0^2(z)}{4\mu_0}, \\[10pt]
        &\quad A_0'(z) ={\frac{p''(z)}{p'(z)\alpha}\left(-\frac{p'(z)}{2K}\right)}^\frac{1}{n}\left(\frac{R^\alpha(z)}{n}+R_0^\alpha(z)\right) - \left(-\frac{p'(z)}{2K}\right)^{\frac{1}{n}}R^{\alpha-1}\theta + \frac{p''(z)R_0^2(z)}{4\mu_0};
    \end{align}

    \text{for} $r > R_0(z)$,
    \begin{align}
        &\quad  f_{0}(z) = R_0(z) v_{r,LSR}(R_0,z)+ R_0^2(z) \left(-\frac{p'(z)}{2K}\right)^\frac{1}{n} \left[ \frac{p''(z)}{p'(z)\alpha n} \left( \frac{R^\alpha(z)}{2} - \frac{R_0^\alpha(z)}{\alpha+2}\right) - \frac{R^{\alpha-1}(z)\theta}{2}\right] . \label{eq:PWM-MSR expressions conical f0(z)}
    \end{align}
  \end{fleqn}
  \item
  High-shear rate (HSR) flow, $\left|\dot{\gamma}_{wall}(z)\right|>\dot{\gamma}_\infty , \ \left|\tau_{wall}(z)\right|>\tau_\infty$,

		for $r \le R_0(z)$,
		\begin{fleqn}[\parindent]
			\begin{align}
			&\quad A_0(z) =\frac{1}{\alpha}\left(-\frac{p'(z)}{2K}\right)^\frac{1}{n}\left(R_\infty^\alpha(z)-R_0^\alpha(z)\right) - \frac{p'(z)}{4}\left(\frac{R_0^2(z)}{\mu_0}+\frac{R^2(z)-R_\infty^2(z)}{\mu_\infty}\right) , \\[10pt]
			&\quad A_0'(z) = -\frac{ p''(z) }{ p'(z)\alpha } \left( -\frac{ p'(z) }{ 2K } \right)^\frac{1}{n} \left(R_\infty^\alpha(z) - R_0^\alpha(z)\right)
			- \frac{ p''(z) }{ 4 } 	\left[ \frac{ R^2(z) + R_\infty^2(z) }{\mu_\infty} 	- \frac{R_0^2(z)}{\mu_0}\right] + \frac{R(z)\theta p'(z)}{2\mu_\infty};
			\end{align}
			\text{for} $ R_0(z) < r \le R_\infty (z) $,
			\begin{align}
			&\quad A_1(z)=-\frac{p'(z)}{4\mu_\infty}\left(R^2(z)-R_\infty^2(z)\right)+\frac{1}{\alpha}\left(-\frac{p'(z)}{2K}\right)^\frac{1}{n}R_\infty^\alpha(z) , \\[10pt]
			&\quad A_1'(z)=-\frac{p''(z)}{4\mu_\infty}\left(R_\infty(z)^2+R^2(z)\right) - \frac{p''(z)}{p'(z) \alpha}\left(-\frac{p'(z)}{2K}\right)^\frac{1}{n} R_\infty^\alpha(z) + \frac{p'(z)R(z)\theta}{2\mu_\infty} , \\[10pt]
			&\quad f_{0}(z) = R_0(z) v_{r,LSR}(R_0,z) - \frac{p''(z)}{p'(z) n \alpha \beta} \left( -\frac{p'(z)}{2K}\right)^\frac{1}{n} R_0^{\alpha+2}(z) + \frac{A_1'(z)R_0^2(z)}{2} \label{eq:PWM-HSR expressions conical f0(z)} ;
			\end{align}
			\text{for} $ r > R_\infty (z)$,
			\begin{align}
			&\quad f_{1}(z) = R_\infty(z) v_{r,MSR}(R_\infty,z) - \frac{ p''(z)R^2_\infty(z) \left( 2R^2(z) - R_\infty^2(z)\right)}{16 \mu_\infty} + \frac{ R_\infty^2(z)\theta p'(z) R(z)}{4 \mu_\infty} . \label{eq:PWM-HSR expressions conical f1(z)}
			\end{align}
		\end{fleqn}
\end{itemize}


\section{Cylindrical flow of a SRB fluid}
For a cylindrical pipe (i.e. corresponding to a conical pipe with $\theta=0^{\circ}$,
$R_{out}=R_{in}=R$, and $\chi=1$) the comparison of pressure gradients in
Eqs. \eqref{eq:conical cond newto 2},\eqref{eq:conical cond powlaw 2} leads to
identically null radial terms, and to a pressure solution as a function of
the axial coordinate $z$ alone.
Therefore, considering the flow rate Eqs. \eqref{eq:flow rate eq con newto},\eqref{eq:flow rate eq con powlaw},
the axial pressure gradient is constant along the nozzle axis and turns out
$-dp/dz=\Delta p/L$.
For a given axial pressure gradient value, the two radius functions $R_0$ and
$R_\infty$ in Eqs. \eqref{eq:R0 and R infinity con} result constant along the axis,
and the quasi-analytical solution in Eqs. \eqref{eq:PWM-LSR expressions conical}-\eqref{eq:PWM-HSR expressions conical}
read

\begin{itemize}
		\item 
		Low-shear rate (LSR) flow, $\left|\dot{\gamma}_{wall}\right| \le \dot{\gamma}_0 ,\ \left|\tau_{wall}\right|\le\tau_0$,
		\begin{subequations}
	    \begin{align}
		&v_{z,LSR}\left(r\right) = \frac{\Delta p R^2}{4L \mu_0}\left[1-\left(\frac{r}{R}\right)^2\right], \qquad \dot{\gamma}_{zr,LSR}\left(r\right) = -\frac{\Delta p}{2L\mu_0}r ;
		\end{align}
	    \end{subequations}
	    with a flow rate $Q = Q_{LSR}$.

		\item 
		Medium-shear rate (MSR) flow, $\dot{\gamma}_0<\left|\dot{\gamma}_{wall}\right|\le\dot{\gamma}_\infty, \tau_0<\left|\tau_{wall}\right|\le\tau_\infty$,
		
	    for $r \le R_0$, 
		\begin{subequations}
			\begin{align}
			&v_{z,LSR}\left(r\right) = A_0-\frac{\Delta p}{4L\mu_0}r^2,  \qquad \dot{\gamma}_{zr,LSR}\left(r\right) = -\frac{\Delta p}{2L\mu_0}r ,\\[10pt]
			&A_0 ={\frac{1}{\alpha}\left(\frac{\Delta p}{2LK}\right)}^\frac{1}{n}\left(R^\alpha-R_0^\alpha\right)\ +\frac{\Delta p R_0^2}{4L\mu_0}, \\[10pt]
			& Q_{LSR} = \pi R_0^2A_0-\frac{\Delta p\pi R_0^4}{8L\mu_0}; \\[10pt]
			\end{align}
	    \text{for} $r > R_0$, 	    
	    	\begin{align}
	    	&v_{z,MSR}\left(r\right) = \left(\frac{\Delta p}{2LK}\right)^\frac{1}{n} \frac{R^\alpha-r^\alpha}{\alpha}, \qquad \dot{\gamma}_{zr,MSR}\left(r\right)= -\left(\frac{\Delta p}{2LK}\right)^\frac{1}{n}r^{1/n},\\[10pt]
            &Q_{MSR}=\frac{2\pi}{\alpha}\left(\frac{\Delta p}{2LK}\right)^\frac{1}{n}\left[\frac{R^\alpha \left(R^2 -R_0^2 \right)}{2}-\frac{R^\beta - R_0^\beta }{\beta}\right] ;
	    	\end{align}	
	    \end{subequations}
	    with a flow rate $Q = Q_{LSR} + Q_{MSR}$.

		\item 
		High-shear rate (HSR) flow, $\left|\dot{\gamma}_{wall}\right|>\dot{\gamma}_\infty , \ \left|\tau_{wall}\right|>\tau_\infty$,
		
		for $r \le R_0$, 
		\begin{subequations}
			\begin{align}
			& v_{z,LSR}\left(r\right) = A_0 - \frac{\Delta p}{4L\mu_0}r^2, \qquad \dot{\gamma}_{zr,LSR}\left(r\right) = -\frac{\Delta p}{2L\mu_0}r, \\[10pt] 
			& A_0 =\frac{1}{\alpha}\left(\frac{\Delta p}{2LK}\right)^\frac{1}{n}\left(R_\infty^\alpha-R_0^\alpha\right) + \frac{\Delta p}{4L}\left(\frac{R_0^2}{\mu_0}+\frac{R^2-R_\infty^2}{\mu_\infty}\right) , \\[10pt]
            & Q_{LSR} = \pi R_0^2 A_0 -\frac{\Delta p\pi R_0^4}{8L\mu_0}; \\[10pt] 
			\end{align}
			\text{for} $ R_0 < r \le R_\infty  $, 	    
			\begin{align}
			& v_{z,MSR}\left(r\right)= -\frac{1}{\alpha}\left(\frac{\Delta p}{2LK}\right)^\frac{1}{n}r^\alpha+A_1, \qquad		
			\dot{\gamma}_{zr,MSR}\left(r\right)=  -\left(\frac{\Delta p }{2LK}\right)^\frac{1}{n}r^\frac{1}{n}, \\[10pt]
			& A_1=\frac{\Delta p}{4L\mu_\infty}\left(R^2-R_\infty^2\right)+\frac{1}{\alpha}\left(\frac{\Delta p}{2LK}\right)^\frac{1}{n}R_\infty^\alpha , \\[10pt]
            & Q_{MSR}=A_1\pi\left(R_\infty^2-R_0^2\right)-\frac{2\pi}{\alpha\beta}\left(\frac{\Delta p}{2LK}\right)^\frac{1}{n}\left(R_\infty^\beta-R_0^\beta \right); \\[10pt]
			\end{align}	
			\text{for} $ r > R_\infty $, 	    
			\begin{align}
			& v_{z,HSR}\left(r\right)= \frac{\Delta p}{4L\mu_\infty}\left(R^2-r^2\right), 
			\qquad \dot{\gamma}_{zr,HSR}\left(r\right)= -\frac{\Delta p}{2L\mu_\infty}r, \\[10pt] 
			& Q_{HSR}=\frac{\Delta p\pi}{8L\mu_\infty}\left(R^2-R_\infty^2\right)^2 ;
			\end{align}
		\end{subequations}
		with a flow rate $Q = Q_{LSR} + Q_{MSR} + Q_{HSR}$.
	
    \end{itemize}
    where $\alpha=(n+1)/n$, and $\beta=(3n+1)/n$.

\end{document}